%% file: main.tex
\documentclass{article}

\usepackage{Definitions/arxiv}

\usepackage[utf8]{inputenc} 
\usepackage[T1]{fontenc}    
\usepackage{url}            
\usepackage{booktabs}       
\usepackage{nicefrac}       
\usepackage{microtype}      

\usepackage{amsmath,amssymb,amsfonts}
\usepackage{algorithmic}
\usepackage{graphicx}
\usepackage{textcomp}

\usepackage{bbm}
\usepackage{bm}
\usepackage{subcaption}
\usepackage{multirow}
\usepackage{array}
\usepackage{float}

\newcommand{\proposal}{\textbf{FAURAS}}
\newcommand{\tableRowSpacing}{1.2pt}

\title{FAURAS: A Proxy-based Framework for Ensuring the Fairness of Adaptive Video Streaming over HTTP/2 Server Push}

\author{
Chanh Minh Tran\\
Graduate School of Engineering and Science\\
Shibaura Institute of Technology\\
\texttt{ma18502@shibaura-it.ac.jp}\\
\And
Tho Nguyen Duc\\
Graduate School of Engineering and Science\\
Shibaura Institute of Technology\\
\texttt{mg18502@shibaura-it.ac.jp}\\
\And
Phan Xuan Tan\\
Department of Information and Communications Engineering\\
Shibaura Institute of Technology\\
\texttt{tanpx@shibaura-it.ac.jp}\\
\And
Eiji Kamioka\\
Graduate School of Engineering and Science\\
Shibaura Institute of Technology\\
\texttt{kamioka@shibaura-it.ac.jp}\\
}

\begin{document}
\maketitle

\begin{abstract}
HTTP/2 video streaming has caught a lot of attentions in the development of multimedia technologies over the last few years.
In HTTP/2, the server push mechanism allows the server to deliver more video segments to the client within a single request in order to deal with the requests explosion problem.
As a result, recent research efforts have been focusing on utilizing such a feature to enhance the streaming experience while reducing the request-related overhead.
However, current works only optimize the performance of a single client, without necessary concerns of possible influences on other clients in the same network.
When multiple streaming clients compete for a shared bandwidth in HTTP/1.1, they are likely to suffer from unfairness, which is defined as the inequality in their bitrate selections.
For HTTP/1.1, existing works have proven that the network-assisted solutions are effective in solving the unfairness problem.
However, the feasibility of utilizing such an approach for the HTTP/2 server push has not been investigated.
Therefore, in this paper, a novel proxy-based framework is proposed to overcome the unfairness problem in adaptive streaming over HTTP/2 with the server push.
Experimental results confirm the outperformance of the proposed framework in ensuring the fairness, assisting the clients to avoid rebuffering events and lower bitrate degradation amplitude, while maintaining the mechanism of the server push feature.
\end{abstract}



  \section{Introduction}\label{section:intro}
  \input{sections/1_introduction.tex}
  
  \section{Related Work}\label{section:relatedWork}
  \input{sections/2_related_work.tex}

  \section{The FAURAS Framework}\label{section:proposal}
  \input{sections/3_proposal.tex}
  
  \section{Performance Evaluation}\label{section:evaluation}
  \input{sections/4_evaluation.tex}
  \section{Discussion}\label{section:discussion}
  \input{sections/5_discussion.tex}
  
  \section{Conclusion and Future Work}\label{section:conclusion}
  \input{sections/6_conclusion.tex}

  \bibliographystyle{unsrt}  
  \bibliography{references.bib}

\end{document}

%% file: sections/1_introduction.tex
Online video streaming and downloads have been consuming a dramatic share of the global network over the last decade and are estimated to reach 82\% of all consumer Internet traffic by 2022 \cite{CiscoReport}.
Such a tremendous growth has put the pressure on the streaming providers to enhance their services in order to serve the viewers with the best quality that corresponds well with the network availability.
To this manner, the HTTP Adaptive Streaming (HAS) has been introduced.
In HAS, each video is encoded in multiple qualities in terms of bitrates; the encoded video with each quality is then chunked into multiple segments with fixed duration (usually 2 to 10 seconds).
At the client side, an adaptive bitrate selection algorithm (ABR) is deployed to continuously measure the current network state (e.g., available bandwidth). Based on the measuring state, the video segments with the suitable quality versions will be fetched and stored into its play-out buffer.
The video starts playing once the buffer is stored at a specified level.
This procedure is repeated throughout the whole streaming session.
The main advantage of this heuristic is that it promotes the client's ability to adapt the video quality to the change of network condition.
As a result, video impairments such as rebuffering events and quality variation can be avoided, thus maintaining a high quality of experience (QoE) for each user.

However, current implementations of HAS mainly utilize the HTTP/1.1 protocol, whose limitations have been exploited in recent studies \cite{http2methodslive, http2livesupershort}.
As the client usually measures the network condition and decides the quality version once a segment is successfully downloaded, a large segment duration obviously decreases the measurement frequency, resulting in slow adaptability that may eventually lead to video impairments in highly unstable network environment. 
Although the aforementioned problem can easily be solved by simply reducing the segment duration (e.g., 1 second or milliseconds), this negatively correlates with the number of segments that need to be delivered (e.g., reducing the segment duration by half means that the number of segments is doubled). 
Due to the pull-based characteristic of HTTP/1.1 - the client has to send a request for every single segment, such an attempt causes the requests explosion that introduces huge overheads to the network infrastructure \cite{lowlatencylivehttp2, costeffectivehttp2}.
Moreover, the number of round-trip time (RTTs) for each request-response pair is also increased, which may degrade the link utilization in critical network conditions \cite{DashHttp2,EvalPushHTTP2}.

Recently, the HTTP/2 protocol has been standardized \cite{rfchttp2} and is now supported by major internet browsers \cite{caniusehttp2}.
The HTTP/2 presents a new feature called \textit{server push}, which allows a server to push responses without having to wait for explicit requests from the client. 
As a result, current researches have focused on leveraging the server push feature of HTTP/2 for adaptive streaming technology \cite{Survey_QoECentricHAS,asurveybitrateadapt}.
Specifically, the so-called \textit{$k$-push} strategy has been adopted, where the client receives $k$ video segments with one request \cite{lowlatencylivehttp2}. 
Prior studies have confirmed the promising performance with short segment duration of such a strategy in reducing request-related overheads \cite{costeffectivehttp2, SeamlessHttp2, DASH2M}, startup and delivering delay \cite{lowlatencylivehttp2,http2livesupershort}, unnecessary RTTs \cite{http2livesupershort,KPushHEVC}, power consumption \cite{powerefficienthttp2,DASH2M} and in improving the QoE \cite{QoEDrivenAdaptiveK,requestadaptation}.
However, those existing works only focused on optimizing the performance of one client without considering the impact on other clients sharing the same bandwidth.
Little attention has been paid to such a mechanism under the multi-clients scenario where the bandwidth competition occurs.

When multiple clients stream a video under a shared bandwidth, the \textit{unfairness} in bitrate selection among the clients can happen in both HTTP/1.1 \cite{whathappen,anexperimental} and HTTP/2 server push \cite{vsip}.
In HTTP/1.1, various research efforts have been made to deal with the unfairness problem, whose deployments varied across different network entities, that is, client-based \cite{FESTIVE,PANDA,QABR,ABMA,BufferBasedHuang}, server-based \cite{ServerShaping,TrackerAssisted,FeedbackControl, Presto}, or network-assisted (e.g., controller, base station, proxy, etc.) \cite{ProxyShapingBandwidth,ShapingHomeGateway,ProxyProactive,designnetworkassisted,QoEProxyFairLTE}.
It has been proven that network-assisted solutions perform best in guiding the clients to select bitrates with respect to fairness, thanks to the ability of in-network entities to globally observe the condition of every client in under its management \cite{Survey_QoECentricHAS, asurveybitrateadapt}.
Although such studies are crucial, to the best of our knowledge, no existing work has aimed at solving the unfairness issue of HAS when employing the HTTP/2 with the server push.

In this paper, we investigate the feasibility of utilizing a network-assisted approach to adaptive streaming over HTTP/2 server push.
A novel framework for \underline{FA}ir and p\underline{U}sh-enabled p\underline{R}oxy-based \underline{A}daptive \underline{S}treaming over HTTP/2 - the \proposal{} framework - is proposed to tackle the unfairness problem when multiple clients compete for a shared bandwidth.
Specifically, by fairly allocating an explicit bandwidth for every client, \proposal{} can effectively eliminate the bandwidth competition, thus avoiding the unfairness problem.
Furthermore, previous studies usually do not consider the situation that clients start the streaming sessions at different time instants, causing abrupt changes of the fair bandwidth that can negatively influence the streaming experience of on-playing clients.
The proposed \proposal{} successfully overcomes this drawback by proactively rewrites the clients' bitrate requests on-the-fly, considering a QoE-aware strategy to harmonize the needs of uninterrupted playback and low bitrate degradation amplitude.
On the other hand, we also found that existing network-assisted solutions in HTTP/1.1 fail to apply in HTTP/2 server push as they tend to drop the pushed segments when bitrate requests are overwritten, thus inefficiently making use of the feature's advantages.
Meanwhile, our proposed framework is able to maintain the mechanism of the HTTP/2 server push by informing the client of the bitrate modification in advance via HTTP headers.
Finally, the experimental results strongly demonstrate that the proposed \proposal{} outperforms the existing works in many different criteria. 
The distinguished contributions of this paper are as follows:
\begin{itemize}

    \item A novel \proposal{} framework is proposed which is the first ever network-assisted approach in adaptive streaming over HTTP/2 with server push. 
    The \proposal{} successfully solves the unfairness problem in bitrate selection among the clients, while also assisting the clients to avoid rebuffering events and to minimize bitrate degradation amplitude.
    
    \item
    The use of the pushed segments are strictly guaranteed, therefore maintaining the mechanism and fully embracing the advantages of the HTTP/2 server push.
    
    \item
    A detailed experimental evaluation is conducted to confirm the superior performance of \proposal{} over the existing methods across various aspects.
    
\end{itemize}

The remainder of the paper is organized as follows:
Section \ref{section:relatedWork} provides an overview of existing investigations and solutions to the unfairness issue in HTTP/1.1.
The proposed \proposal{} framework for solving the unfairness in HTTP/2 server push is presented in details and evaluated in Section \ref{section:proposal} and \ref{section:evaluation}, respectively.
Section \ref{section:discussion} discusses the effectiveness of the proposed framework.
Finally, Section \ref{section:conclusion} concludes this paper.

%% file: sections/2_related_work.tex
The behavior of multiple clients sharing a fixed bottleneck bandwidth in adaptive streaming over HTTP/1.1 was first investigated in \cite{anexperimental}.
A performance evaluation of major commercial and open source adaptive streaming players with two competing clients was conducted, showing that the unfairness in bitrate selection occurred.
The authors argued that the issue was not related to the well-known congestion control mechanism of TCP but to the mismatch of bandwidth estimations among the clients.
This finding was confirmed and investigated further in \cite{whathappen}.
It was discovered that, due to the temporal overlap of the clients' segment download states, which were called the \textit{ON-OFF periods}, a client might overestimate its available bandwidth, thus selecting a too high bitrate version and unfairly occupying the utilization for other clients.
As a result, several attempts have been carried out to solve the unfairness in adaptive streaming over HTTP/1.1.

Existing solutions to the unfairness of HAS in HTTP/1.1 can be categorized into client-based, server-based and network-assisted methods \cite{asurveybitrateadapt}.
Client-based adaptation schemes \cite{FESTIVE,PANDA,ABMA,BufferBasedHuang,QABR} make use of the computational power of clients' devices to adapt the video bitrate based on measurements of different adaptation metrics, such as available bandwidth, playback buffer, instantaneous QoE, etc.
Although benefiting the system deployment and scalability, such methods usually perform suboptimally due to insufficient information regarding the entire network conditions \cite{FINEAS}.
On the other hand, server-based solutions \cite{ServerShaping,TrackerAssisted,FeedbackControl,Presto} utilize the advantages of centralized video servers to perform overall bitrate optimization and exchange insights about the clients' statuses.
However, these solutions either limit the system scalability when the number of clients increases or introduce high overhead and complexity that may harm the network efficiency \cite{asurveybitrateadapt,Survey_QoECentricHAS}.
Meanwhile, network-assisted approaches \cite{ProxyShapingBandwidth,ShapingHomeGateway,ProxyProactive,designnetworkassisted,QoEProxyFairLTE} employ in-network entities (e.g., network edges, controllers, proxies, etc.) to assist the clients' bitrate decisions.
Thanks to the general observation of the network, such approaches have shown a high performance in both small and large networks.
A bandwidth shaping mechanism was deployed at the residential gateway in \cite{ProxyShapingBandwidth,ShapingHomeGateway} to assign fair bandwidth slices to the clients.
As such, the bandwidth competition was avoided and the bitrate difference among the clients were minimized.
In \cite{ProxyProactive}, the authors discussed the use of a network proxy in overwriting the client's bitrate request to meet with the calculated fair share.
The work in \cite{designnetworkassisted} compared the performance of different Software-Defined Network based (SDN-based) methods, including bandwidth shaping, bitrate guidance and hybrid method, found out that all methods provided remarkable improvement of fairness.
Likewise, several proxy-based methods were evaluated in \cite{QoEProxyFairLTE} considering the fairness in terms of user's QoE under the Long-Term Evolution (LTE) network.
The aforementioned researches are crucial for the mass adoption of the adaptive streaming technology.
However, to the best of our knowledge, no existing work has ever considered the unfairness problem when utilizing the HTTP/2 server push.

The existence of the unfairness in adaptive streaming over HTTP/2 server push has been confirmed in our previous work \cite{vsip}.
Therefore, in this paper, we set light for solving such a problem with a network-assisted approach by proposing the \proposal{} framework.


%% file: sections/3_proposal.tex
\begin{figure}[ht]
  \centering
    \includegraphics[width=1.0\linewidth]{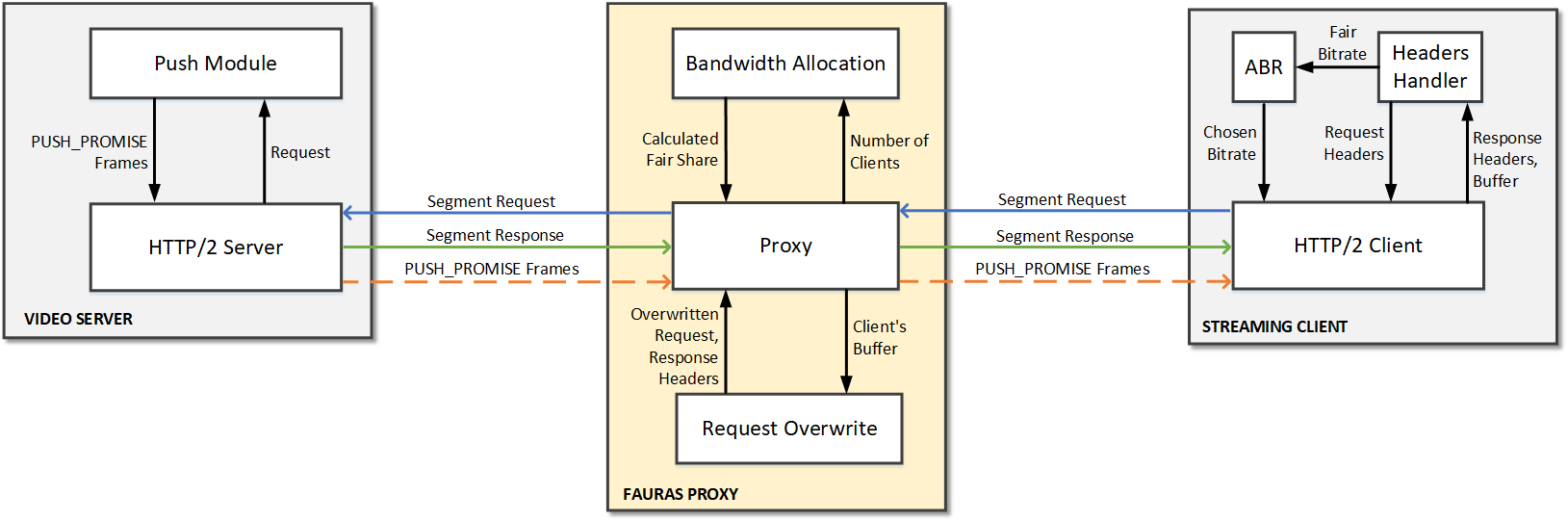}
  \caption{Block diagram of the \proposal{} framework}
  \label{fig:3_Overview}
\end{figure}

Figure \ref{fig:3_Overview} depicts the block diagram of the proposed \proposal{} framework. 
In this framework, requests from multiple clients to the video server are routed through a proxy which is located in-between (e.g., at network edges).
At the proxy, as soon as a client joins and starts its streaming session, a fair bandwidth is calculated and assigned separately to each client presenting in the network.
This corresponds with the function of the \textbf{Bandwidth Allocation} module.
The \textbf{Request Overwrite} module, on the other hand, is responsible for rewriting the segment's request into one with the bitrate matching the fair share.
The overwriting decision of this module is based on a QoE-aware strategy that takes into account the QoE-related metrics, that is, rebuffering event and bitrate degradation amplitude.
In addition, when the \textbf{Request Overwrite} module modifies a bitrate request, a notification message is signaled to the corresponding client via HTTP headers so that the client will not discard the upcoming pushed segments.
The whole procedure of \proposal{} is kept transparent so that it does not require significant modifications for both the server and client side.
In the following subsections, the function of each module is presented and discussed in details.
Some notations and definitions used in this paper are summarized in table \ref{tbl:3_Notations}.

\input{tables/3_Notations.tex}

  
  \subsection{Bandwidth Allocation}
  \input{sections/3_proposal/3-1_BandwidthAllocation.tex}
  
  \subsection{Request Overwrite}
  \input{sections/3_proposal/3-2_RequestOverwrite.tex}

%% file: tables/3_Notations.tex
\begin{table}[ht]
  \caption{Notations and definitions used in this paper}
  \centering
  \begin{tabular}{ccl}
    \toprule
    \textbf{Notation} & \textbf{Unit} & \textbf{Definition} \\
    \midrule
    $L$ & second & the segment duration \\[\tableRowSpacing]
    $t$ & second & a time instant within the streaming session \\[\tableRowSpacing]
    $X_t$ & & the total number of streaming clients currently presenting in the network at time $t$ \\[\tableRowSpacing]
    $a_x$ & & a streaming client currently presenting in the network, $x \in [1,X_t]$ \\[\tableRowSpacing]
    \midrule
    $B_t$ & second  & the actual buffer level at time $t$ \\[\tableRowSpacing]
    $B^e_t$ & second & the estimated buffer level at time $t$ \\[\tableRowSpacing]
    $B_{max}$ & second & the maximum buffer level \\[\tableRowSpacing]
    \midrule
    $N$ & & the total number of available bitrate versions \\[\tableRowSpacing]
    $\Re$ &  & the set of available bitrate versions, $\Re = \{R^0, R^1, ..., R^N\}$ \\[\tableRowSpacing]
    $R^f_t$ & kbps & the calculated fair bitrate at time $t$ \\[\tableRowSpacing]
    $r_{x,t}$ & kbps & the requested bitrate of client $a_x$ at time $t$ \\[\tableRowSpacing]
    \midrule
    $S^{n}_i$ &  & the segment at index $i^{th}$ with bitrate $R^n$ \\[\tableRowSpacing]
    $k$ &  & the number of segments received in a push cycle \\[\tableRowSpacing]
    \midrule
    $\mathbb{C}$ & kbps & the maximum available bandwidth \\[\tableRowSpacing]
    $C^f_t$ & kbps & the calculated fair bandwidth at time $t$ \\[\tableRowSpacing]
    $c_{x,t}$ & kbps & the bandwidth allocated for client $a_x$ at time $t$ \\[\tableRowSpacing]
    \bottomrule
  \end{tabular}
  \label{tbl:3_Notations}
\end{table}

%% file: sections/3_proposal/3-1_BandwidthAllocation.tex
\subsubsection{Determining the Fair Share}
  \input{sections/3_proposal/3-1_BandwidthAllocation/3-1-1_DefiningFair.tex}
  
\subsubsection{Problem when New Clients Join}
  \label{subsec:NewClientsJoin}
  \input{sections/3_proposal/3-1_BandwidthAllocation/3-1-2_ProblemNumClientsChange.tex}

%% file: sections/3_proposal/3-1_BandwidthAllocation/3-1-1_DefiningFair.tex

A client starts its streaming session in the \textit{buffering state} where it aggressively downloads video segments to fill the play-out buffer.
Once the buffer reaches its maximum at $B_{max}$ seconds, the client switches to the \textit{steady state} when it plays out and downloads segments simultaneously to maintain the buffer stable at $B_{max}$.
Suppose that at time $t$ in the steady state, client $a_x$ initiates a new push cycle with the bitrate $r_{x,t}=R^n$ and sends request for the next $k$ segments $\{S^{n}_{i},S^{n}_{i+1}...S^{n}_{i+k-1}\}$, where the first segment $S^{n}_{i}$ is called the \textit{lead segment}.
As the video keeps playing during the download of the segments, the buffer will lose an amount corresponding to the downloading time, which can be approximated as $k*L*\frac{r_{x,t}}{c_{x,t}}$.
After that, $k*L$ seconds more will be added to the buffer.
Therefore, when the client finishes the push cycle, the estimated buffer condition at time $t'$ can be determined by the Eq. (\ref{eqn:Buffer}):

\begin{equation}
    \label{eqn:Buffer}
    B^e_{t'} = B_t + k*L - k*L*\frac{r_{x,t}}{c_{x,t}} \leq B_{max}
\end{equation}

Since the proxy has a global view of all HAS traffics under its management, it is designated for dividing a suitable bandwidth slice for every of its client.
Such an approach, as discussed in Section \ref{section:relatedWork}, can effectively eliminate the bandwidth competition.
The \textbf{Bandwidth Allocation} module calculates the fair share by dividing the available bandwidth by the number of currently-streaming clients, which is described by the Eq. (\ref{eqn:FairBW}):

\begin{equation}
  C_t^f = \frac{\mathbb{C}}{X_t}
  \label{eqn:FairBW}
\end{equation}

where $C_t^f$ demonstrate the fair share at time $t$.
At the client side, the deployed ABR decides the bitrate to be requested based on the bandwidth $c_{x,t}$ it observes.
In general, a bitrate decision should always satisfy:

\begin{equation}
\label{eqn:FairBR}
\begin{split}
  r_{x,t} &\leq R_t^f \\
  \text{subject to \:} R_t^f &= max\{R^n\in\Re | R^n \leq C_t^f\}
\end{split}
\end{equation}

where $R_t^f$ is the fair bitrate version corresponding with the fair bandwidth $C_t^f$ at time $t$.
Inferring from Eq. (\ref{eqn:Buffer}), a bitrate higher than $R_t^f$ will probably harm the play-out buffer as the buffer has to spend more time to download the segments than the duration it is added.
Due to the fact that maximizing the QoE is among the crucial goals of every ABR \cite{asurveybitrateadapt,Survey_QoE,Survey_QoECentricHAS}, it obviously tends to decide the bitrate that is equal to $R_t^f$ for its client since this is the best possible video quality under the bandwidth $C_t^f$.
Consequently, as all clients are assigned with the same bandwidth $C_t^f$ by the \textbf{Bandwidth Allocation} module, they are likely to select similar video bitrates, thus achieving fairness.
It should be noted that, we assume every client shares the same characteristics (e.g., screen size, device type, subscription plan, etc.) in this paper.

%% file: sections/3_proposal/3-1_BandwidthAllocation/3-1-2_ProblemNumClientsChange.tex
\begin{figure}[h]
  \centering
    \includegraphics[width=0.8\linewidth]{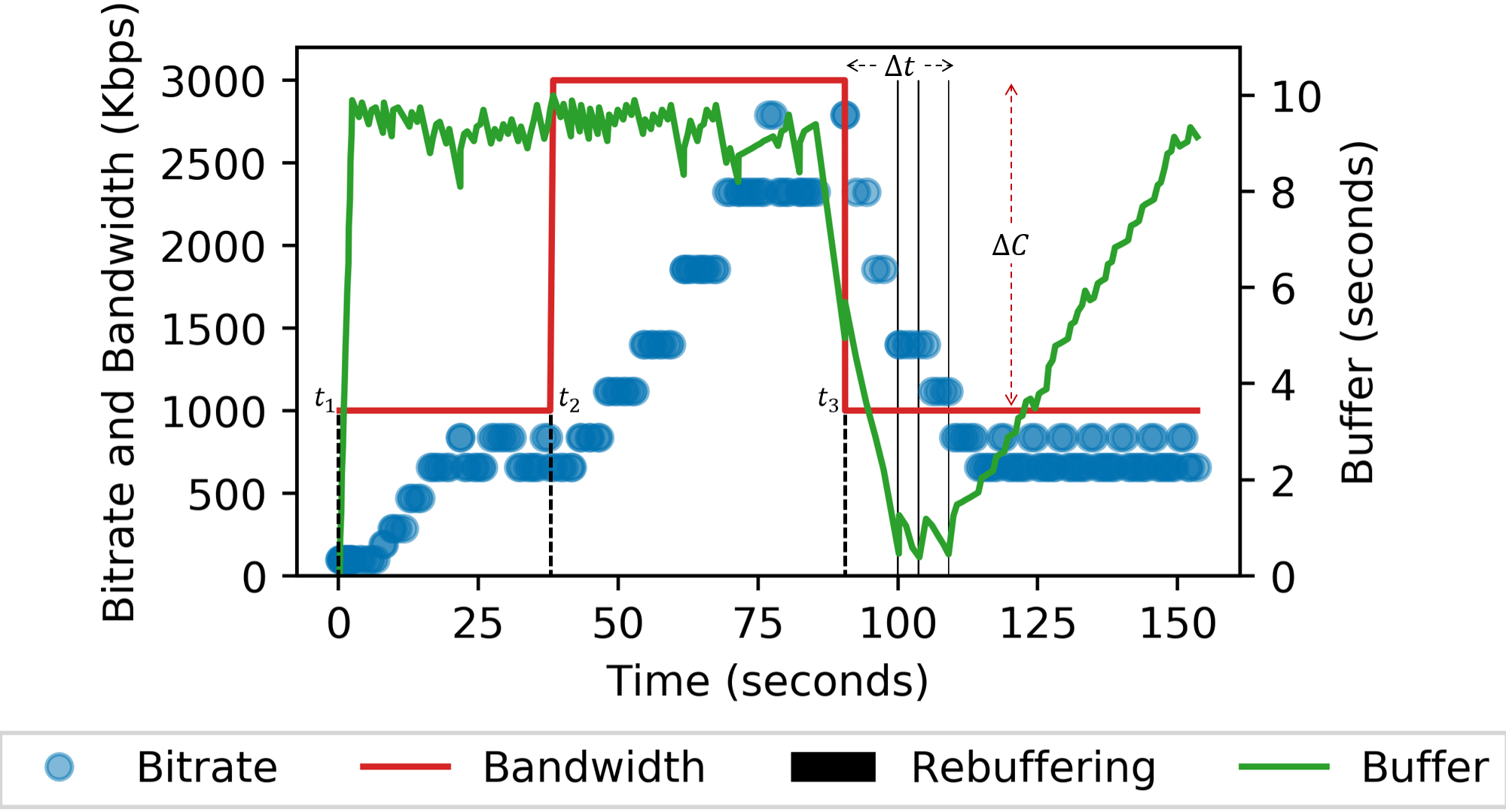}
  \caption{An example experiment of how a client reacts to the changes of fair share}
  \label{fig:3-1_BandwidthAllocation}
\end{figure}

Every time existing clients stop or new clients start their streaming sessions, the module recalculates and assigns the new fair share $C_{t'}^f$ for all active clients.
Figure \ref{fig:3-1_BandwidthAllocation} illustrates an example experiment of the two cases.
In this experiment, one client is tested by streaming 200 segments from a video server via a proxy.
The detailed settings will be presented later in Section \ref{section:evaluation}.
To simulate the start/stop of clients, we set the proxy to allocate different bandwidth at different periods for the client as described in Table \ref{tbl:3_BandwidthAllocation}.

\input{tables/3_BandwidthAllocation}

The client starts with the initial bandwidth of 1 Mbps at time $t_1$.
After downloading segment $60^{th}$ at time $t_2$, the bandwidth increases to 3 Mbps.
This simulates the case when existing clients finish their streaming sessions and leave the network.
As the number of clients has decreased ($X_{t_2}<X_{t_1}$), the bandwidth assigned for each client is increased ($C_{t_2}^f>C_{t_1}^f$), which also increases the fair bitrate ($R_{t_2}^f\geq{}R_{t_1}^f$). 
Such a situation deals no negative impact but, instead, can increase the user's satisfaction.

On the other hand, from segment $121^{th}$ at time $t_3$, the bandwidth is shaped to only 1 Mbps, resembling that new clients have joined to the network ($X_{t_3}>X_{t_2}$).
At this time, the current client will experience a drop of bandwidth ($C_{t_3}^f<C_{t_2}^f$), which is the main consideration in the development of HAS.
As a result, it has to decrease its bitrate to match with the new fair bitrate ($R_{t_3}^f \leq R_{t_2}^f$).
Given the fact that most ABRs often decrease the bitrate step-by-step to maintain the user's QoE \cite{GradualBitrateSwitch,SeamlessHttp2}, the client usually needs $\Delta{}t$ seconds to adapt to the new condition and the buffer is likely to degrade during the time.
As a consequence, a large gap $\Delta{}C$ between $C_{t_3}^f$ and $C_{t_2}^f$ will doubtlessly result in a large adaptation delay $\Delta{}t$.
Despite that such a large $\Delta{}t$ can help the client experience the high bitrate longer, it is likely to increase the risk of buffer underflow.
This is illustrated clearly in Fig. \ref{fig:3-1_BandwidthAllocation}: the client's buffer depletes 3 times during this period.
This has shown that, under such a situation, the \textbf{Bandwidth Allocation} module itself cannot efficiently help the client react in time to bandwidth degradation.
To overcome this disadvantage, in the next subsection, a \textbf{Request Overwrite} module, which is the most important component of our system, is presented to assist clients shifting to the fair bitrate before draining out their buffers.

%% file: tables/3_BandwidthAllocation.tex

\begin{table}[ht]
  \caption{Bandwidth allocated to the client at specific periods in the streaming session}
  \centering
  \begin{tabular}{cccc}
    \toprule
    \textbf{Segments} & 1 - 60 & 61 - 120 & 121 - 200 \\
    \midrule
    \textbf{Bandwidth (Mbps)} & 1 & 3 & 1 \\
    \bottomrule
  \end{tabular}
  \label{tbl:3_BandwidthAllocation}
\end{table}

%% file: sections/3_proposal/3-2_RequestOverwrite.tex
\subsubsection{Overwriting Strategy}
  \input{sections/3_proposal/3-2_RequestOverwrite/3-2-1_Trigger.tex}
  
\subsubsection{Handling the Server Push Mechanism}
\label{subsection:handlingpush}
  \input{sections/3_proposal/3-2_RequestOverwrite/3-2-2_HandlingPush.tex}

%% file: sections/3_proposal/3-2_RequestOverwrite/3-2-1_Trigger.tex
The \textbf{Request Overwrite} module is responsible for proactively rewriting the bitrate of the segment request to the one matching with the fair bandwidth if it detects that the client is demanding a higher value.
The operation is performed on-the-fly and no additional communication link is required for both the server and the client side.
A safe and aggressive strategy is to immediately overwrite the client's request once the bitrate exceeds the fair video rate.
This effectively reduces the adaptation delay, hence maintaining the buffer at a high level and definitely preventing the rebuffering from happening. 
However, such an approach results in a large down-switching amplitude $\Delta{}r$ (i.e., the bitrate difference between the current bitrate decision and one right before it), which has been proven to have negative impact on the user's QoE \cite{Survey_QoE}.
Therefore, it is necessary to develop an overwriting strategy that can minimize the bitrate difference while ensuring a smooth video playback.

In the \proposal{} framework, we propose a QoE-aware strategy for overwriting the bitrate request.
The video will not be rebuffered if there is at least one segment in the buffer \cite{BufferBasedHuang}.
Considering the case of the HTTP/2 server push, during a push cycle, the client fetches $k$ segments with the same bitrate which cannot be changed in-between.
To this manner, for the k-push strategy, it is optimal to maintain $k$ segments in the buffer at all time.
Thus, the \textbf{Request Overwrite} module will correct the client's bitrate decision once the following condition is met:

\begin{equation}
\label{eqn:Trigger}
\left\{\begin{matrix}
\begin{aligned}
r_{x,t} &> R_t^f
\\
B^e_{t'} &< k*L  
\end{aligned}
\end{matrix}\right.
\end{equation}

When the client sends a request for segment $S^n_i$, it will notify the proxy of its current buffer condition $B_t$.
Such information is delivered via an explicit HTTP header included to the request so that no extra connection is created, thus limiting unnecessary overheads \cite{FINEAS}.
If the requested bitrate $r_{x,t}$ exceeds the fair value, the proxy extracts the buffer information from the request headers to estimate the buffer $B^e_{t'}$ after finishing the download of the push cycle.
Once $B^e_{t'}$ is expected to drop under the duration of $k$ segments, the proxy will overwrite the bitrate decision to $R^f_t$.

%% file: sections/3_proposal/3-2_RequestOverwrite/3-2-2_HandlingPush.tex

\input{figures/3-3-2_flows.tex}

In the HTTP/2 server push, the pushed resources are stored in the browser's push cache of the client for later use \cite{PushToCache1,powerefficienthttp2}.
Figure \ref{fig:3-3-2_flows} depicts examples of this procedure in HAS; Fig. \ref{fig:3-3-2_flow_a} describes a typical scenario that the proxy only forwards the client's segment request, while rewriting it into different bitrate in Fig. \ref{fig:3-3-2_flow_b}.
Assuming that in a 2-push mechanism (i.e., for 1 request, 2 segments are delivered), at time $t$, a client $a_x$ decides the bitrate $r_{x,t} = R^n$ for the next push cycle and prepares to send the request for segment $S^{n}_{i}$.
The client first checks its push cache and finds that $S^{n}_{i}$ hasn't been downloaded.
Therefore, it sends the request to the video server, which traverses via a proxy.

In Fig. \ref{fig:3-3-2_flow_a}, the proxy keeps the request unmodified and forwards it to the server. 
The server sends the requested segment $S^{n}_{i}$ to the client's buffer via an HTTP response and pushes segment $S^{n}_{i+1}$ to the push cache via a PUSH\_PROMISE frame.
Then, as the client is still in the push cycle, its bitrate decision remains unchanged at $R^{n}$.
However, before sending the request for $S^{n}_{i+1}$, the client sees that such a segment is already in the push cache (whole or partly).
Accordingly, the client only needs to take this segment out and directly add into the play-out buffer.
This procedure is repeated until the end of the streaming session.

In the case of Fig. \ref{fig:3-3-2_flow_b}, the proxy rewrites the bitrate of $S^{n}_{i}$ into $R^{n'}$.
Therefore, $S^{n'}_{i}$ and $S^{n'}_{i+1}$ are delivered into the client's buffer and push cache, respectively.
Due to the fact that the client is not aware of its bitrate decision being modified, we speculate that it still expects the bitrate $R^{n}$ for the next segment.
Thus, when checking the push cache, the client cannot see $S^{n}_{i+1}$ as the server pushed $S^{n'}_{i+1}$ instead.
As a result, the client discards $S^{n'}_{i+1}$ and sends request for $S^{n}_{i+1}$, which completely compromises the mechanism of the server push.
Hence, not only does the client fail to make use of the advantages discussed in Section \ref{section:relatedWork}, but also it wastes the server's network resources utilized for the push segment \cite{WastePush}.

An experiment has been conducted in order to prove the hypothesized phenomenon. 
We set up one single client streaming 100 video segments from the server via a proxy, under an unlimited bandwidth and with the 2-push strategy.
At the client side, a simple algorithm is deployed that it only selects the lowest bitrate $R^0$ for the whole session.
In this experiment, the proxy overwrites all segment requests from the client to ones with the second-lowest bitrate $R^1$.
The detailed settings will be presented in Section \ref{section:evaluation}.
Table \ref{tbl:3_Push} summarizes the total number of responses ($Res$) and PUSH\_PROMISE frames ($PP$) captured on the client's device, in comparison with the cases when the overwrite function is not activated.

\input{tables/3_Push}

As for HTTP/1.1, it is obvious that the client had to send requests for all 100 segments due to the pull-based characteristic.
For the case of the 2-push strategy, the client received 2 segments for each request sent (i.e., 1 in the response to the request and 1 delivered via the PUSH\_PROMISE frame).
Thus, the number of requests was reduced to half that corresponds to the increase of PUSH\_PROMISE frames. 
However, when rewriting the requests, not only did the number of requests remain similar as in HTTP/1.1, but also the number of PUSH\_PROMISE frames was double.
This confirms the above hypothesis: the pushed segments will be discarded if the overwritten bitrate does not meet the client's desire.

To overcome this issue, an essential notification function is proposed in the \textbf{Request Overwrite} module.
Everytime a bitrate request is modified, the module explicitly informs its client by simply adding an HTTP header to the response of the lead segment.
Once the client receives such an information, it is forced to select the bitrate for the remaining segments in the push cycle as one the proxy requires.
Therefore, although the overwrite function can work with any ABR, clients do need a minor modification to support extracting the information in the headers and modifying its bitrate decision to match with the proxy's demand.

%% file: figures/3-3-2_flows.tex
\begin{figure}[ht!]
  \centering
  \begin{subfigure}{0.9\linewidth}
    \centering
    \includegraphics[width=\textwidth]{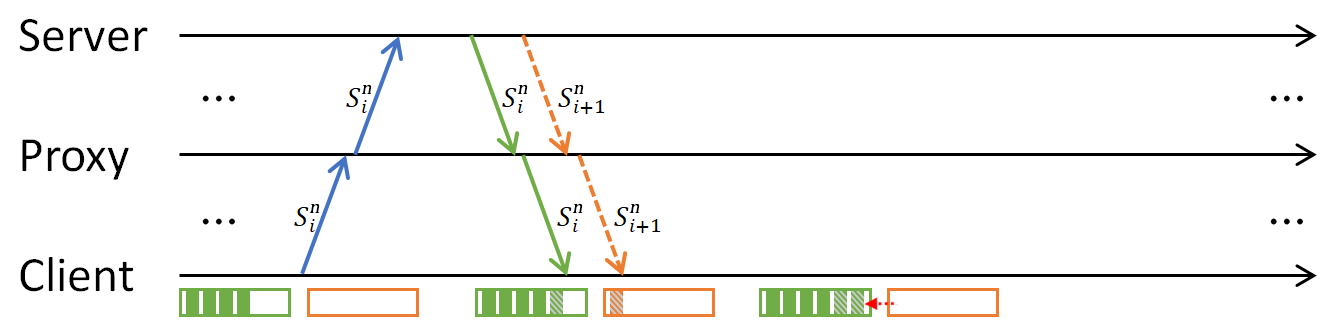}
    \caption{Proxy only forwards the client's request}
    \label{fig:3-3-2_flow_a}
  \end{subfigure}
  
  \vspace{6pt}
  
  \begin{subfigure}{0.9\linewidth}
    \centering
    \includegraphics[width=\textwidth]{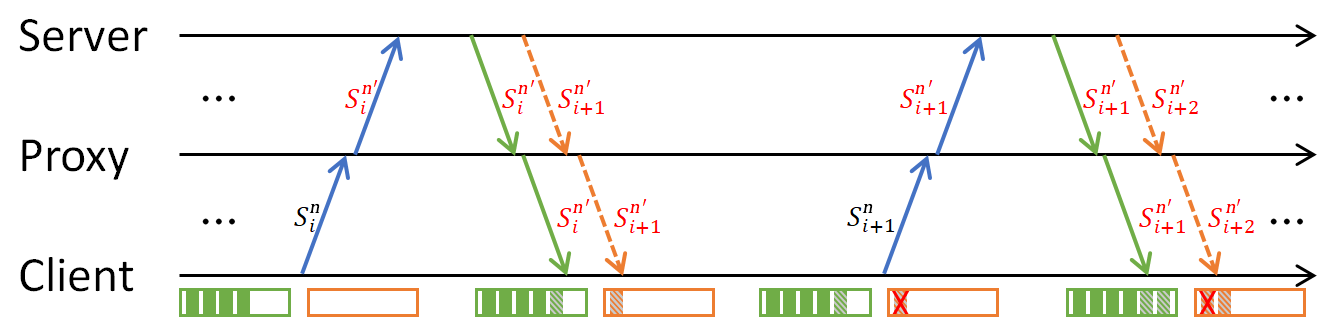}
    \caption{Proxy overwrites the client's request}
    \label{fig:3-3-2_flow_b}
  \end{subfigure}
  
  \vspace{9pt}
  
  \begin{subfigure}{0.85\linewidth}
    \centering
    \includegraphics[width=\textwidth]{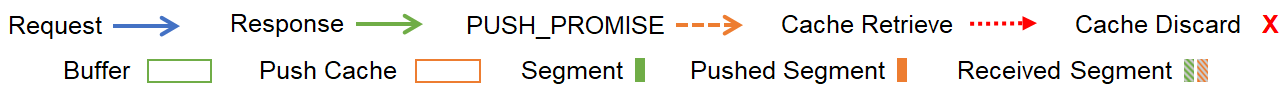}
    \label{fig:3-3-2_legend}
  \end{subfigure}
  
  \caption{Requests/responses flows in adaptive streaming over HTTP/2 with server push when (a) the proxy only forwards the client's request and (b) the proxy overwrites the client's request}
  \label{fig:3-3-2_flows}
\end{figure}

%% file: tables/3_Push.tex
\begin{table}[ht]
  \caption{Total number of requests and PUSH\_PROMISE frames capturing on the client's device}
  \centering
  \begin{tabular}{cccc}
    \toprule
    \textbf{Metric} & \textbf{HTTP/1.1} & \textbf{HTTP/2 2-push} & \textbf{HTTP/2 2-push with Overwrite}\\
    \midrule
    $Res$ & 100 & 50 & 100 \\
    $PP$ & 0 & 50 & 99\\ 
    \bottomrule
  \end{tabular}
  \label{tbl:3_Push}
\end{table}

%% file: sections/4_evaluation.tex
\subsection{Experimental Setup}
\label{subsection:4-1}
  \input{sections/4_evaluation/4-1_Implementation.tex}
 
\subsection{Evaluation Scenarios and Metrics}
\label{4_scenario-metrics}
  \input{sections/4_evaluation/4-2_Method.tex}
  
\subsection{Detailed Results}
  \input{sections/4_evaluation/4-3_Results.tex}

%% file: sections/4_evaluation/4-1_Implementation.tex
\begin{figure}[h]
  \centering
    \includegraphics[width=0.6\linewidth]{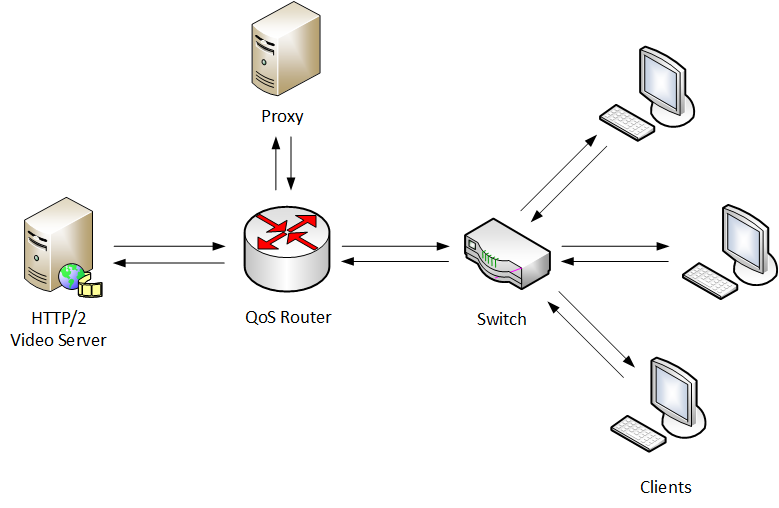}
  \caption{Experiment topology}
  \label{fig:4-1_topology}
\end{figure}

In this subsection, we describe our experimental settings for evaluating the proposed \proposal{}.
Our testbed consists of an HTTP/2 server, a proxy and multiple HTTP/2 streaming clients as shown in Fig. \ref{fig:4-1_topology}.
The settings of the server and clients are similar to that of \cite{vsip}.
The server is run on a Windows 10 Core i7 physical machine with 8GB of RAM and is implemented based on \textit{Jetty} \cite{jetty}, while each client is a Core i5 physical machine with 4GB of RAM.
On the server, the \textit{DASH.js} streaming player \cite{dashjs} is packaged as a web application, which will be run at the client side via Chrome web browser.
At the client side, the \textit{Wireshark} \cite{wireshark} is deployed to monitor the number of HTTP responses and PUSH\_PROMISE frames the clients receive.

On the same machine as the server, an Ubuntu 18.04 LTS virtual machine with 3GB of RAM and 3 cores of CPU is installed to implement the proxy and its modules.
A virtual QoS Router is created with the \textit{RYU SDN} framework \cite{ryusdn} for monitoring and controlling the network flows between the server and clients.
All traffic from the clients are transparently routed to the proxy which is deployed based on the \textit{mitmproxy} \cite{mitmproxy}.
Every time a client sends a request for the MPD file to start the streaming session, the proxy recalculates the fair share and signals the QoS Router to assign a bandwidth slice for the new client and adjust the bandwidth given for existing clients.
When the proxy judges that the client's bitrate decision would cause rebuffering, it rewrites the URL in the GET request to one matching the fair bitrate and informs the client by adding a header to the response. 

In our experiment, the server provides the \textit{Big Buck Bunny} video with 11 versions of bitrate, i.e. $\Re = \{99,192,285,470,656,838,1118,1401,1855,2324,2791\}$ (Kbps) as in \cite{vsip}.
Each version is chunked into segments of 1 second.
The maximum bandwidth $\mathbb{C}$ is fixed at 3 Mbps, which is sufficient for a client to reach the highest bitrate.
As recommended in \cite{http2livesupershort}, the $k$-push value in this experiment is set to 2.
At the client side, the maximum buffer level $B_{max}$ is set to 10 seconds.
The \textit{FESTIVE} \cite{FESTIVE} algorithm is chosen as the client's ABR, with minor modifications to support the HTTP/2 server push mechanism and communications via HTTP headers.
The FESTIVE employs the harmonic mean smoothing technique for bandwidth estimation, gradual bitrate transition and randomized segment download scheduler for tackling the unfairness from the client side.
Please note that the main focus of this paper is the performance of network-assisted solutions in solving the unfairness problem of HAS when using the HTTP/2 server push; a comparison of different ABRs will be left as future work.

For performance comparison, in addition to the proposed \proposal{}, the following methods are considered under the same settings as above:
\begin{itemize}
    \item \textbf{No-Proxy:} 
    This represents the typical situation when the proxy is not activated and all clients compete fiercely for the bandwidth.
    \\
    \item \textbf{Reactive:}
    The proxy only allocates the fair bandwidth separately for every client and leaves the client's bitrate decision untouched. This corresponds to the work in \cite{ProxyShapingBandwidth}, which is proposed for HTTP/1.1.
    \\
    \item \textbf{Proactive:}
    An implementation of the method proposed in \cite{ProxyProactive}, which is also for HTTP/1.1. Each client gets its own fair bandwidth given by the proxy. If a client requests a bitrate that is higher than the calculated fair value, the proxy immediately corrects it and forwards to the server.
\end{itemize}

 
  

%% file: sections/4_evaluation/4-2_Method.tex
Table \ref{tbl:4-2_Scenarios} shows the experimental scenarios for assessing the performance of the proposed \proposal{} and reference methods.

\input{tables/4-2_Scenarios}

In this experiment, each client streams a total of 200 segments from the server.
Similar to existing researches, a common scenario that multiple clients join together is considered, denoted as scenario \#1.
Specifically, the mentioned methods are tested on 3 subscenarios \textit{1-A}, \textit{1-B} and \textit{1-C} with the number of clients ranging from 2 to 4, respectively.
In addition, we also evaluate those methods with scenario \#2 when new clients join in the middle of an existing client's session, causing the change of the fair bandwidth.
A single client $a_1$ firstly plays the video alone in the network, utilizing the whole fixed bandwidth for itself.
After it finishes the download of the $100^{th}$ segment, the client $a_2$ joins and introduces a bandwidth gap $\Delta{}C$ of 1.5 Mbps observing from the original client.
This is denoted as subscenario \textit{2-A}.
Another subscenario, denoted as \textit{2-B}, is that 2 clients $a_2$ and $a_3$ join together at the same moment as above, which causes the bandwidth gap to increase to 2 Mbps.

The proposal and aforementioned methods are evaluated in various aspects.
The following metrics are considered in both scenario \#1 and \#2:
\begin{itemize}
    \item 
        Unfairness Index ($F$): 
        Solving the unfairness problem is the main focus of this paper. 
        The unfairness index $F$ across all clients is calculated based on the Jain Fairness index \cite{JainFair} as in Eq. \ref{eqn:F}.
        A lower value of $F$ indicates a better performance.
        The unfairness performance in scenario \#1 is determined by averaging $F$ from the beginning to the end of the streaming session.
        On the other hand, in scenario \#2, it is averaged from the time when the new clients $a_2$ and $a_3$ join to the network to the time when $a_1$ finishes the video. 
    
        \begin{equation}
        \label{eqn:F}
            F = \sqrt{1-\frac{{(\sum_{x=1}^{X_t} r_{x,t})}^2}{X_t*\sum_{x=1}^{X_t}{r_{x,t}}^2}}
        \end{equation}
        \\
    \item
        Number of Rebuffering Events ($Rebuff$): 
        Rebuffering event should be eliminated since it is the most serious video impairment that degrades the user's QoE \cite{futureinternet}.
        In this evaluation, the total number of rebuffering events of all clients is collected.

\end{itemize}



In scenario \#2, the performances of all network-assisted methods are investigated further. 
Specifically, the explicit performance of client $a_1$ is assessed based on the following metrics:

\begin{itemize}
    \item 
        Adaptation Delay ($\Delta{}t$): 
        As discussed in Section \ref{section:proposal}, a large adaptation delay can lead to rebuffering events. 
        Therefore, it is necessary to reduce such delay to ensure a smooth streaming playback. 
        $\Delta{}t$ is calculated from the time when the new clients $a_2$ and $a_3$ join to the time when the client $a_1$ achieves the fair bitrate.
        \\
        
    \item
        Bitrate Degradation Amplitude ($\Delta{}r$):
        Existing researches have found that viewers often react negatively to abruptl and large degradation in visual quality \cite{Survey_QoE}. 
        For this reason, the amplitude of bitrate down-switching should be minimized.
        In this evaluation, $\Delta{}r$ accounts for the gap between the bitrate at the first time it reaches the fair value with the one right before it.
        \\
        
        
    \item
        Number of HTTP Responses ($Res$) and number PUSH\_PROMISE frames ($PP$):
        These metrics are to evaluate the ability of \proposal{} and referenced methods in ensuring the HTTP/2 server push mechanism.
        For streaming 200 segments from the server with the 2-push strategy, baseline values of 100 HTTP responses and 100 PUSH\_PROMISE frames must be precisely achieved.
        
\end{itemize}

A summary of the above evaluation metrics is shown in Table \ref{tbl:4-2_Metrics}.
In the following subsection, the experimental results in each scenario are presented in details.

\input{tables/4-2_Metrics}
 

%% file: tables/4-2_Scenarios.tex
\begin{table}[h]
  \caption{Experimental scenarios}
  \centering
  \begin{tabular}{ccl}
    \toprule
    \textbf{Scenario} & \textbf{Subscenario} & \textbf{Description}\\
    \midrule
    \multirow{3}{6em}{\centering \textbf{Scenario \#1}} &
    \textbf{1-A} & 2 clients join together\\
    & \textbf{1-B} & 3 clients join together\\ 
    & \textbf{1-C} & 4 clients join together\\
    \midrule
    \multirow{2}{6em}{\centering \textbf{Scenario \#2}} & \textbf{2-A} & 1 client $a_1$ joins first, then 1 more client $a_2$ joins later\\
    & \textbf{2-B} & 1 client $a_1$ joins first, then 2 clients $a_2$ and $a_3$ join together later\\ 
    \bottomrule
  \end{tabular}
  \label{tbl:4-2_Scenarios}
\end{table}

%% file: tables/4-2_Metrics.tex
\begin{table}[h]
  \caption{Evaluation Metrics}
  \centering
  \begin{tabular}{cclc}
    \toprule
    \textbf{Metric} & \textbf{Unit} & \textbf{Description} & \textbf{Evaluated Scenario}\\
    \midrule
    $F$ & & the unfairness index among all clients & \#1 \& \#2 
    \\[\tableRowSpacing]
    $Rebuff$ & & the number of rebuffering events of all clients & \#1 \& \#2 
    \\[\tableRowSpacing]
    $\Delta{}t$ & second & the adaptation delay of client $a_1$ & \#2 
    \\[\tableRowSpacing]
    $\Delta{}r$ & Kbps & the bitrate degradation amplitude of client $a_1$ & \#2 
    \\[\tableRowSpacing]
    $Res$ & & the number of HTTP responses the client $a_1$ receives & \#2 
    \\[\tableRowSpacing]
    $PP$ & & the number of PUSH\_PROMISE frames the client $a_1$ receives & \#2
    \\
    \bottomrule
  \end{tabular}
  \label{tbl:4-2_Metrics}
\end{table}

%% file: sections/4_evaluation/4-3_Results.tex

The experiment of each subscenario in the scenarios provided in the previous subsection was conducted 5 times.
The average results are summarized as below.




\subsubsection{Scenario \#1}
\label{subsection:4-3-1_scenario1}
  \input{sections/4_evaluation/4-3_Results/4-3-1_Scenario1}

\subsubsection{Scenario \#2}
\label{subsection:4-3-2_scenario2}
  \input{sections/4_evaluation/4-3_Results/4-3-2_Scenario2}

%% file: sections/4_evaluation/4-3_Results/4-3-1_Scenario1.tex
\input{tables/4_Scenario1}

Table \ref{tbl:4_Scenario1} shows the achieved unfairness index of all methods as well as the total number of rebuffering events occurring in scenario \#1.
It should be noted that, for the number of rebuffering events, the results show an accumulated value of all running times instead of an average value. 
In general, all network-assisted methods provided relatively identical unfairness performances that significantly outperformed the \textbf{No-Proxy} method.
Particularly, the fairness is improved by 4.29 times on average.
The results of number of rebuffering events express the similar trend.
While the proxy-based methods successfully eliminated such an impairment, the \textbf{No-Proxy} method performed more poorly when the number of clients increases.
This is because the more clients there were in the network, the more fiercely they competed for the bandwidth.
In Fig. \ref{fig:4_bitrate}, the time-varying performance of each method is illustrated.
Due to similar behaviors, only the subscenario 1-B is shown.

\input{figures/4_bitrate}

It can be observed from Fig. \ref{fig:4_bitrate_no} that client $a_2$ usually selected a bitrate higher than the fair share (1 Mbps) and consumed the bandwidth of other clients.
For this reason, not only that client $a_1$ and $a_3$ failed to reach the fair bitrate but their play-out buffers also became highly unstable and eventually depleted (as for client $a_3$).
Meanwhile, under assistance of the proxies in the \textbf{Reactive}, \textbf{Proactive} and proposed \proposal{} method (Fig. \ref{fig:4_bitrate_re}, \ref{fig:4_bitrate_proact} and \ref{fig:4_bitrate_proposal}), all clients were able to select similar bitrates and maintained the buffers stable.
In the next part, the performance of the network-assisted solutions are investigated further, with respect to the ability to assist the client to adapt to the changes of the fair bandwidth.

%% file: tables/4_Scenario1.tex
\begin{table}[h]
  \caption{The performance of unfairness index and number of rebuffering events in scenario \#1}
  \centering
  \begin{tabular}{ cccccc}
    \toprule
    \textbf{Metric}& \textbf{Subscenario} &\textbf{No-Proxy} & \textbf{Reactive} & \textbf{Proactive} & \proposal{}\\
    \midrule
    \multirow{3}{5em}{\centering $F$} 
    & \textbf{1-A} & 0.2512 & 0.0304 & 0.0392 & 0.0391\\
    & \textbf{1-B} & 0.2551 & 0.0658 & 0.0670 & 0.0661\\
    & \textbf{1-C} & 0.3879 & 0.1148 & 0.1156 & 0.1133\\
    \midrule
    \multirow{3}{5em}{\centering $Rebuff$} 
    & \textbf{1-A} & 0 & 0 & 0 & 0\\
    & \textbf{1-B} & 3 & 0 & 0 & 0\\
    & \textbf{1-C} & 5 & 0 & 0 & 0\\
    \bottomrule
  \end{tabular}
  \label{tbl:4_Scenario1}
\end{table}

%% file: figures/4_bitrate.tex
\begin{figure}[h!]
  \centering
  \begin{subfigure}{0.47\linewidth}
    \centering
    \includegraphics[width=\textwidth]{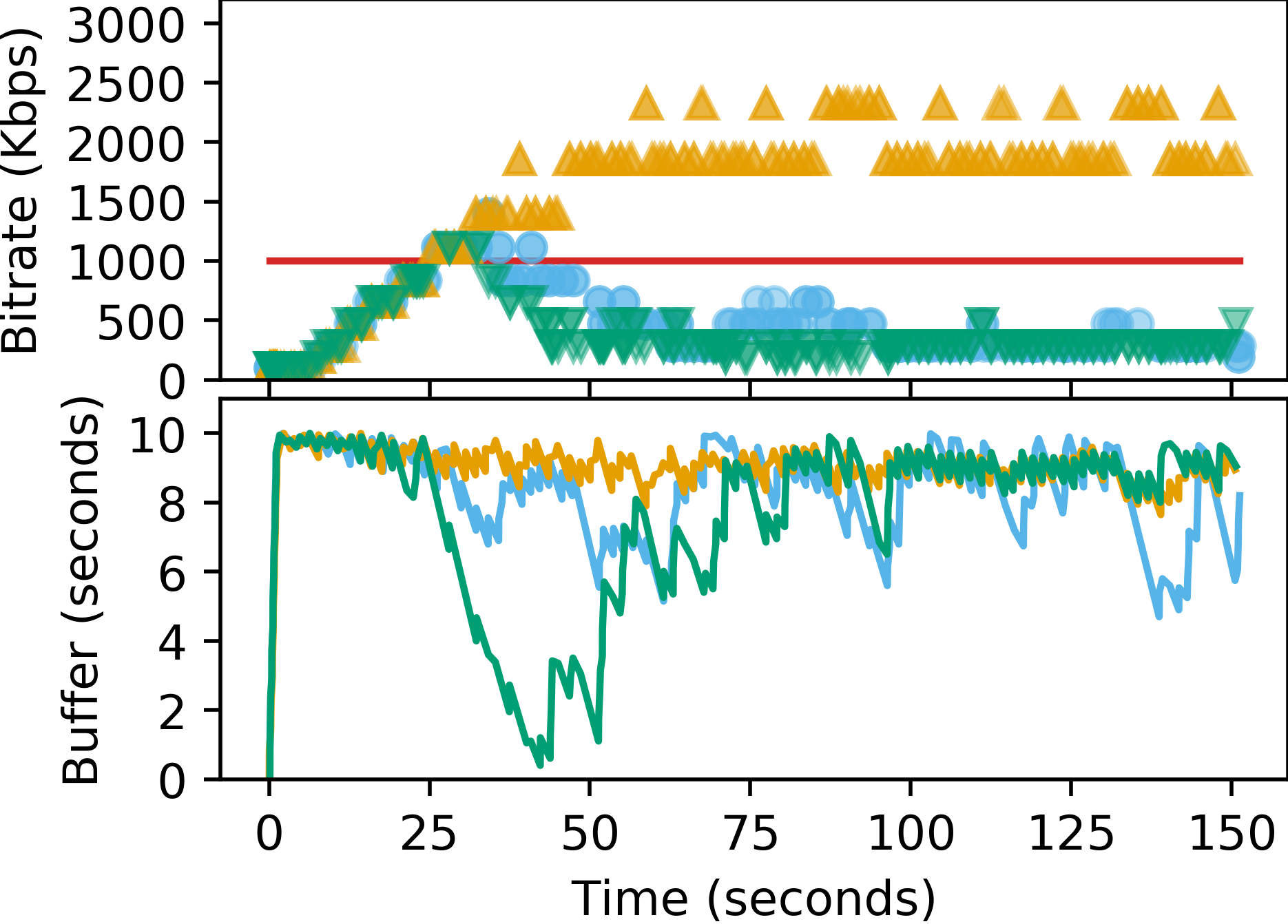}
    \caption{No-Proxy}
    \label{fig:4_bitrate_no}
  \end{subfigure}
  \hfill
  \begin{subfigure}{0.47\linewidth}
    \centering
    \includegraphics[width=\textwidth]{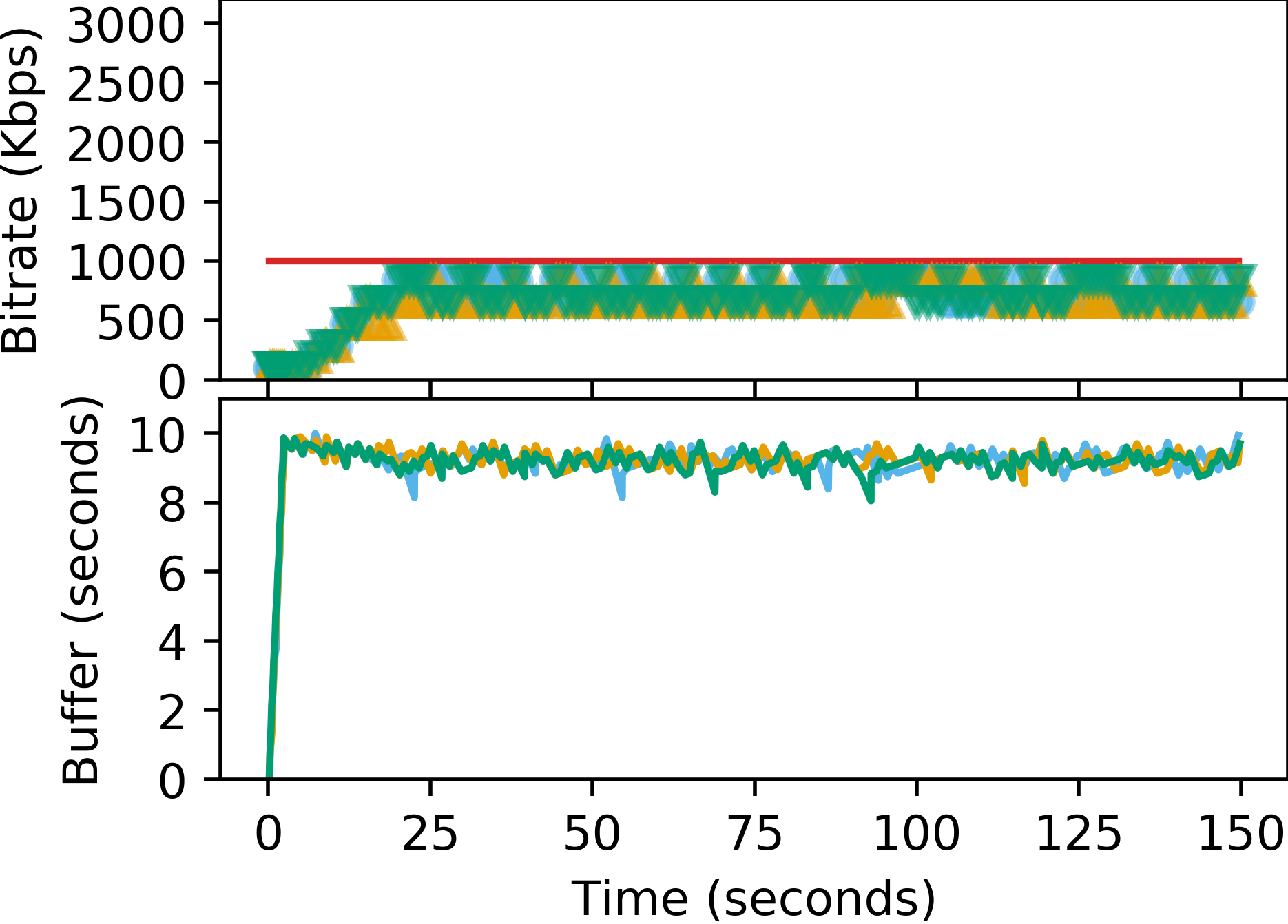}
    \caption{Reactive}
    \label{fig:4_bitrate_re}
  \end{subfigure}
  
  \vspace{6pt}
  
  \begin{subfigure}{0.47\linewidth}
    \centering
    \includegraphics[width=\textwidth]{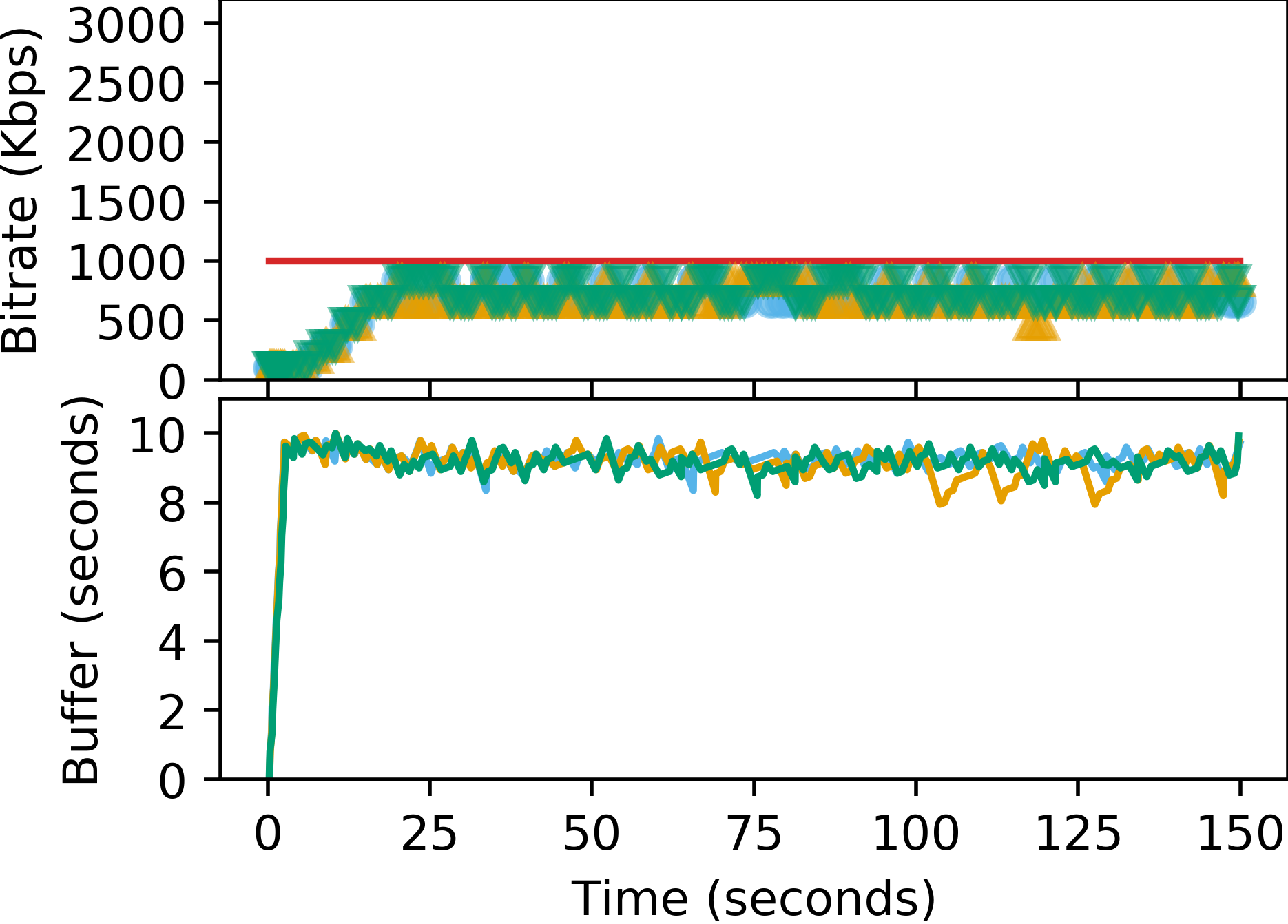}
    \caption{Proactive}
    \label{fig:4_bitrate_proact}
  \end{subfigure}
  \hfill
  \begin{subfigure}{0.47\linewidth}
    \centering
    \includegraphics[width=\textwidth]{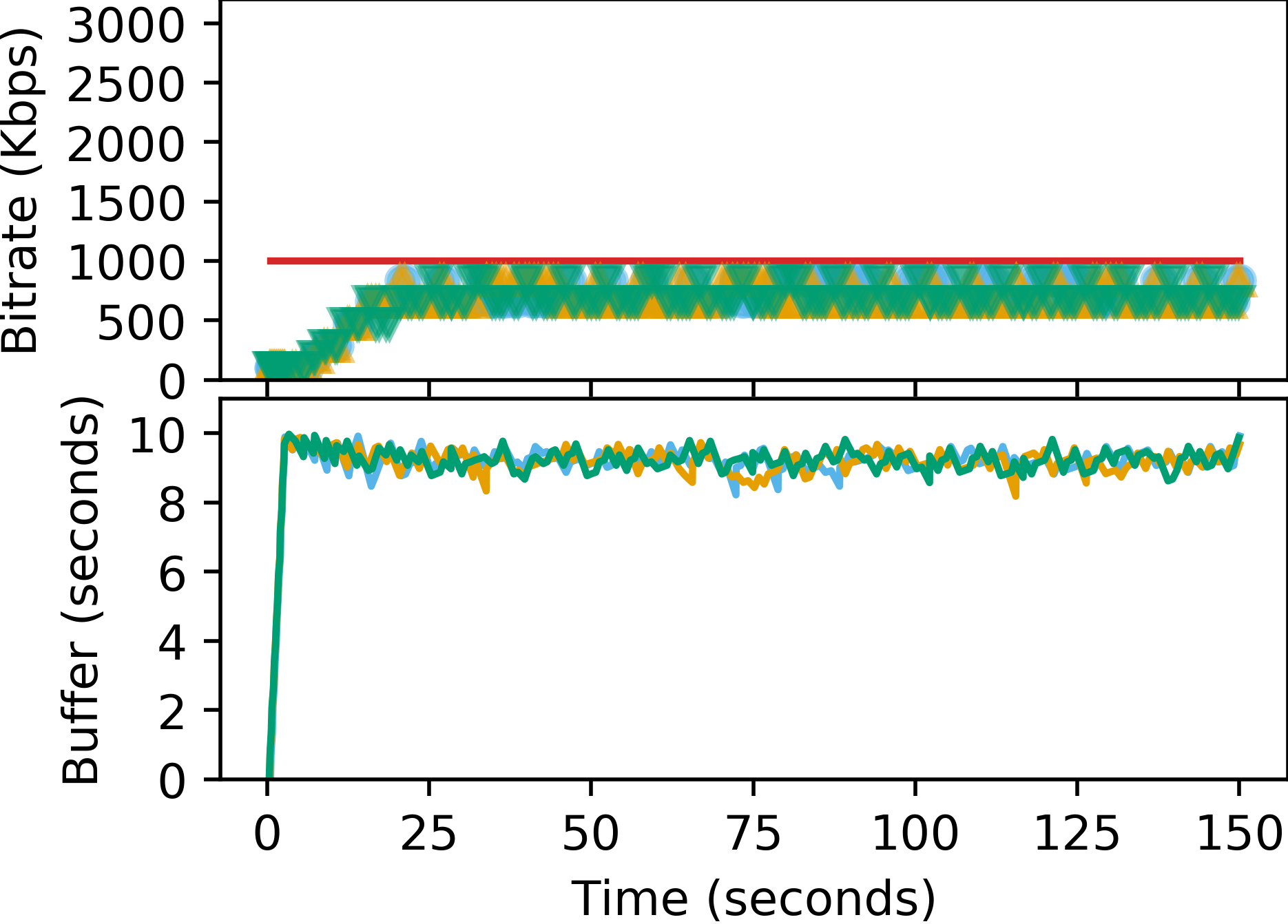}
    \caption{Proposal}
    \label{fig:4_bitrate_proposal}
  \end{subfigure}
  
  \vspace{8pt}
  
  \begin{subfigure}{0.9\linewidth}
    \centering
    \includegraphics[width=0.7\textwidth]{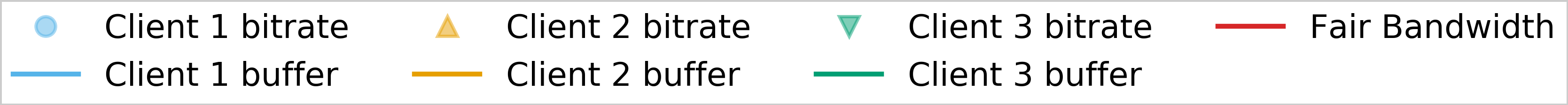}
  \end{subfigure}
  
  \caption{Time-varying performances of all methods in scenario \#1}
  \label{fig:4_bitrate}
\end{figure}

%% file: sections/4_evaluation/4-3_Results/4-3-2_Scenario2.tex
The results of unfairness and the number of rebuffering events in Scenario \#2 are summarized in Table \ref{tbl:4_Scenario2}.
Meanwhile, Fig. \ref{fig:4_bitrate_1-1} and \ref{fig:4_bitrate_1-2} illustrate the time-varying performances of the methods in subscenarios 2-A and 2-B, respectively.

\input{tables/4_Scenario2}

\input{figures/4_bitrate_1-1}
\input{figures/4_bitrate_1-2}

The unfairness performances of the evaluated methods show the similar tendency as in Scenario \#1. 
The solutions with proxy improved the fairness by 1.72 times on average comparing to the \textbf{No-Proxy} method.
It is noticeable that the improvement was less than that in Scenario \#1.
Actually, this is understandable due to the characteristics of the client's ABR.
As shown in Fig. \ref{fig:4_bitrate_1-1} and \ref{fig:4_bitrate_1-2}, the new clients $a_2$ and $a_3$ had to select their bitrate step-by-step from the lowest version when they started their streaming session, while $a_1$ was already at high levels.
As such, the unfairness index performed worse during this period and recovered once all clients achieved the fair bitrate (1401 Kbps for subscenario 2-A and 838 Kbps for subscenario 2-B).

Likewise, the \textbf{No-Proxy} method still underwent rebuffering events, while the proposed \proposal{} and the \textbf{Proactive} method showed consistent efficiency in avoiding them.
However, despite the same performance was achieved in subscenario 2-A, the \textbf{Reactive} method failed to prevent rebuffering events in subscenario 2-B when the number of newly-join clients increased.
In addition, all rebuffering events when using this method only occurred on client $a_1$, who started the streaming session earlier and experienced the bandwidth change when other clients joined.
The numerical statistics of the adaptation delay, along with bitrate degradation amplitude, of the network-assisted methods are shown in Table \ref{tbl:4_QoERelated} in order to explain this underperformance.

\input{tables/4_QoERelated}

The \textbf{Reactive} method only allocated the bandwidth for the client and respected the gradual quality deterioration of the ABR.
As shown in Fig. \ref{fig:4_bitrate_1-1_re} and \ref{fig:4_bitrate_1-2_re}, the video bitrate only dropped by 1 level at a time.
For this reason, this method provided the lowest $\Delta{}r$ among the evaluated methods in both subscenarios.
However, it resulted in the highest $\Delta{}t$ that might cause the client to fail to maintain enough buffer.
For the subscenario 2-A, the fair bandwidth dropped from 3 Mbps to 1.5 Mbps and the fair bitrate had to decrease by 3 levels from 2791 Kbps to 1401 Kbps.
In this case, the client $a_1$ still managed adapt to the new fair share before the buffer ran out (Fig. \ref{fig:4_bitrate_1-1_re}).
However, in subscenario 2-B, the fair bandwidth dropped from 3 Mbps to 1 Mbps that decreased the fair bitrate by 5 levels from 2791 Kbps to 838 Kbps.
Thus, $\Delta{}t$ of the client $a_1$ increased by 1.9 times and caused its buffer to deplete (Fig. \ref{fig:4_bitrate_1-2_re}).

On the other hand, the \textbf{Proactive} method immediately rewrote the client's request to match with the fair bitrate.
Observing from Fig. \ref{fig:4_bitrate_1-1_proact} and Fig. \ref{fig:4_bitrate_1-2_proact}, the client $a_1$ decreased its bitrate straight away instead of switching the bitrate steadily via intermediate levels.
Obviously, the $\Delta{}t$ was reduced to the lowest, hence effectively avoiding the rebuffering events.
As a consequence, this method induced the largest $\Delta{}r$.

Meanwhile, it is shown that the proposed \proposal{} framework harmonized the performance of the other two methods.
For subscenario 2-A, the \proposal{} and the \textbf{Reactive} method performed equally in terms of $\Delta{}r$, which were approximately 2.24 times lower than the \textbf{Proactive} method.
This is visualized by Fig. \ref{fig:4_bitrate_1-1_proposal}, the bitrate of client $a_1$ also decreased gradually as when using the \textbf{Reactive} method.
Therefore, they required similar $\Delta{}t$ values that did not harm the playout buffers.
In the case of subscenario 2-B, the $\Delta{}r$ of the \proposal{} was slightly higher than the \textbf{Reactive} method (1.8 times) but was still significantly lower than the \textbf{Proactive} method (3.5 times).
Inferring from Fig. \ref{fig:4_bitrate_1-2_proposal}, at 86s, the client decided to switch from 1401 Kbps to 838 Kbps and skipped the value at 1118 Kbps.
As a result, the $\Delta{}t$ was reduced comparing to the \textbf{Reactive} method and the buffer was able to recover before fully depleted, thus avoiding rebuffering events.

\begin{figure}[h!]
  \centering
    \includegraphics[width=1\linewidth]{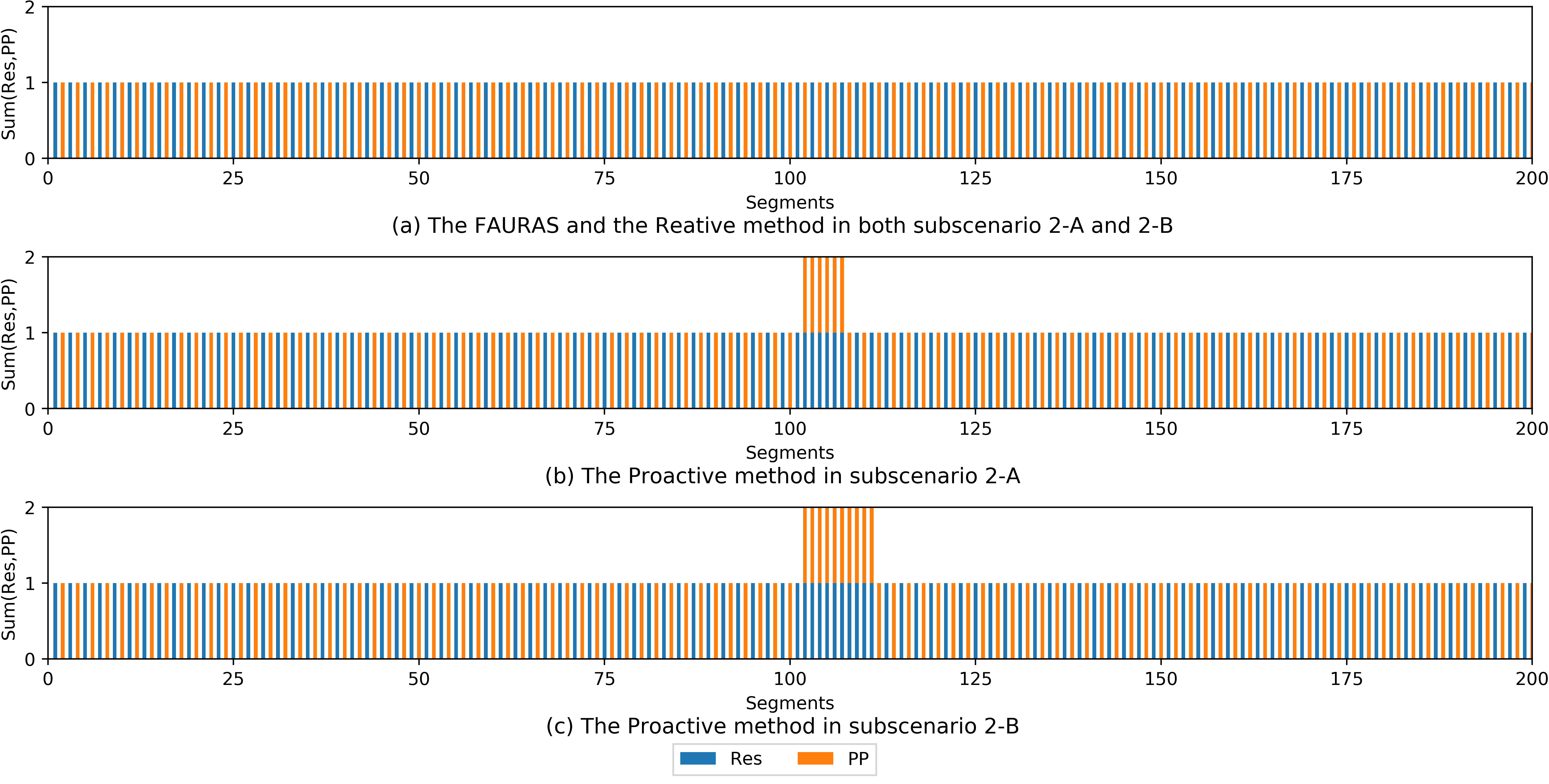}
  \caption{The sum of HTTP Responses and PUSH\_PROMISE frames the client $a_1$ received per segment}
  \label{fig:4_sum_res_pp}
\end{figure}

Additionally, Table \ref{tbl:4_Push} shows the number of HTTP responses and the number of PUSH\_PROMISE frames client $a_1$ received throughout the whole session, while Fig. \ref{fig:4_sum_res_pp} illustrates the sum of those metrics for each video segment specifically.

\input{tables/4_Push}

According to Table \ref{tbl:4_Push}, the \proposal{} and \textbf{Reactive} method show consistently identical statistics with the baseline as described in \ref{4_scenario-metrics}.
The behavior of those methods is represented in Fig. \ref{fig:4_sum_res_pp}a.
It is obvious that the client alternately received either a single HTTP response or a single PUSH\_PROMISE frame for every segment throughout the streaming session.
The \textbf{Proactive} method, however, failed to achieve such a performance,
the $Res$ and $PP$ values exceeded the baseline in both cases.
Referring to Fig. \ref{fig:4_sum_res_pp}b (subscenario 2-A) and \ref{fig:4_sum_res_pp}c (subscenario 2-B), after segment $100^{th}$ which was the moment that the new clients joined, the client $a_1$ consecutively  received both an HTTP responses and a PUSH\_PROMISE frame for 6 and 10 segments, respectively.
This indicates that, although those segments had been pushed by the server, the client still requested and used those in the HTTP responses instead of fetching from its push cache.
As result, those pushed segments were wasted and the client switched back to the pull-based mechanism of HTTP/1.1 during this period.

%% file: tables/4_Scenario2.tex
\begin{table}[ht]
  \caption{The performance of unfairness index and number of rebuffering events in scenario \#2}
  \centering
  \begin{tabular}{ cccccc}
    \toprule
    \textbf{Metric}& \textbf{Subscenario} &\textbf{No-Proxy} & \textbf{Reactive} & \textbf{Proactive} & \proposal{}\\
    \midrule
    \multirow{2}{5em}{\centering $F$} 
    & \textbf{2-A} & 0.4021 & 0.2250 & 0.2204 & 0.2267\\
    & \textbf{2-B} & 0.3910 & 0.2460 & 0.2190 & 0.2429\\
    \midrule
    \multirow{2}{5em}{\centering $Rebuff$} 
    & \textbf{2-A} & 0 & 0 & 0 & 0\\
    & \textbf{2-B} & 5 & 9 & 0 & 0\\
    \bottomrule
  \end{tabular}
  \label{tbl:4_Scenario2}
\end{table}

%% file: figures/4_bitrate_1-1.tex
\begin{figure}[ht!]
  \centering
  \begin{subfigure}{0.47\linewidth}
    \centering
    \includegraphics[width=\textwidth]{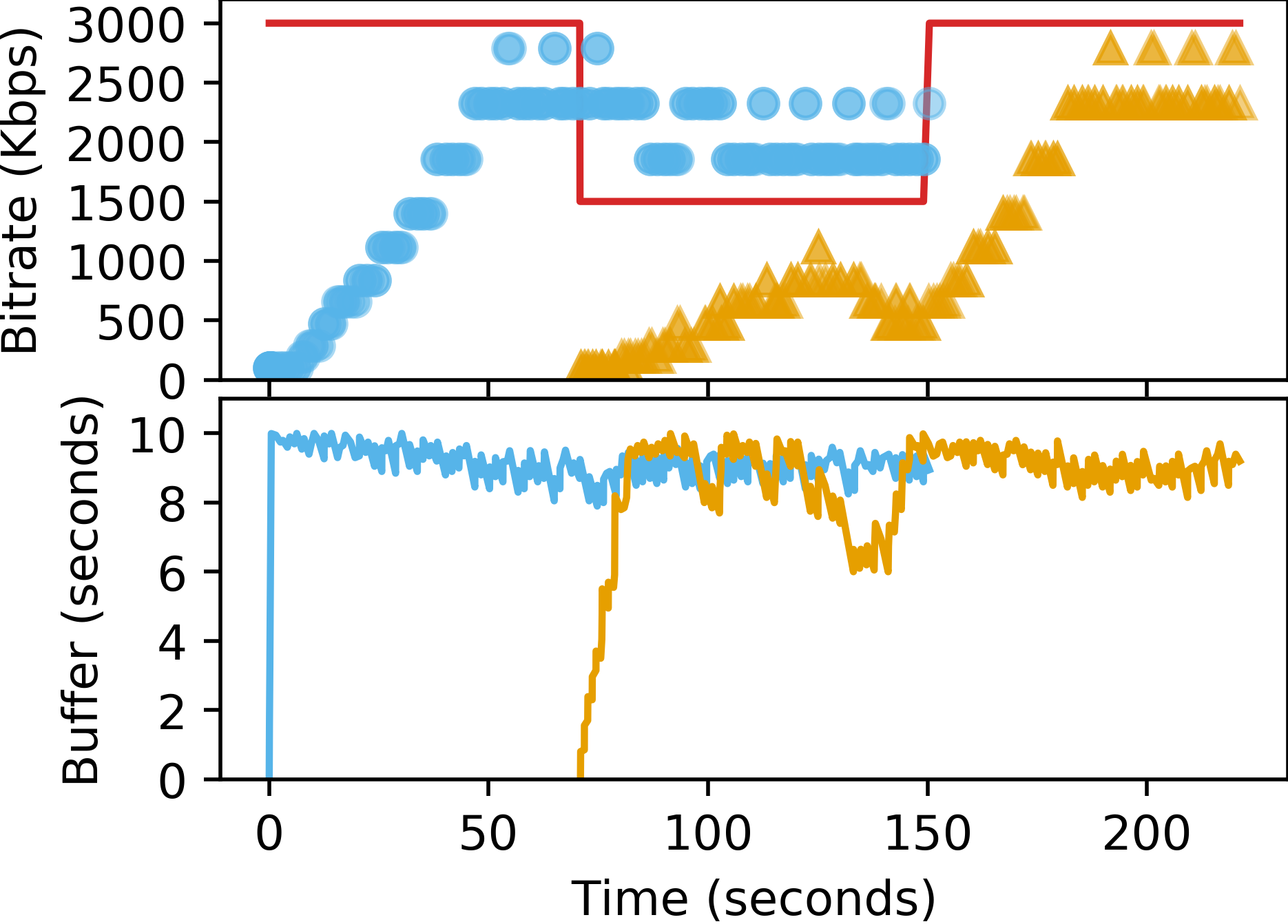}
    \caption{No-Proxy}
    \label{fig:4_bitrate_1-1_no}
  \end{subfigure}
  \hfill
  \begin{subfigure}{0.47\linewidth}
    \centering
    \includegraphics[width=\textwidth]{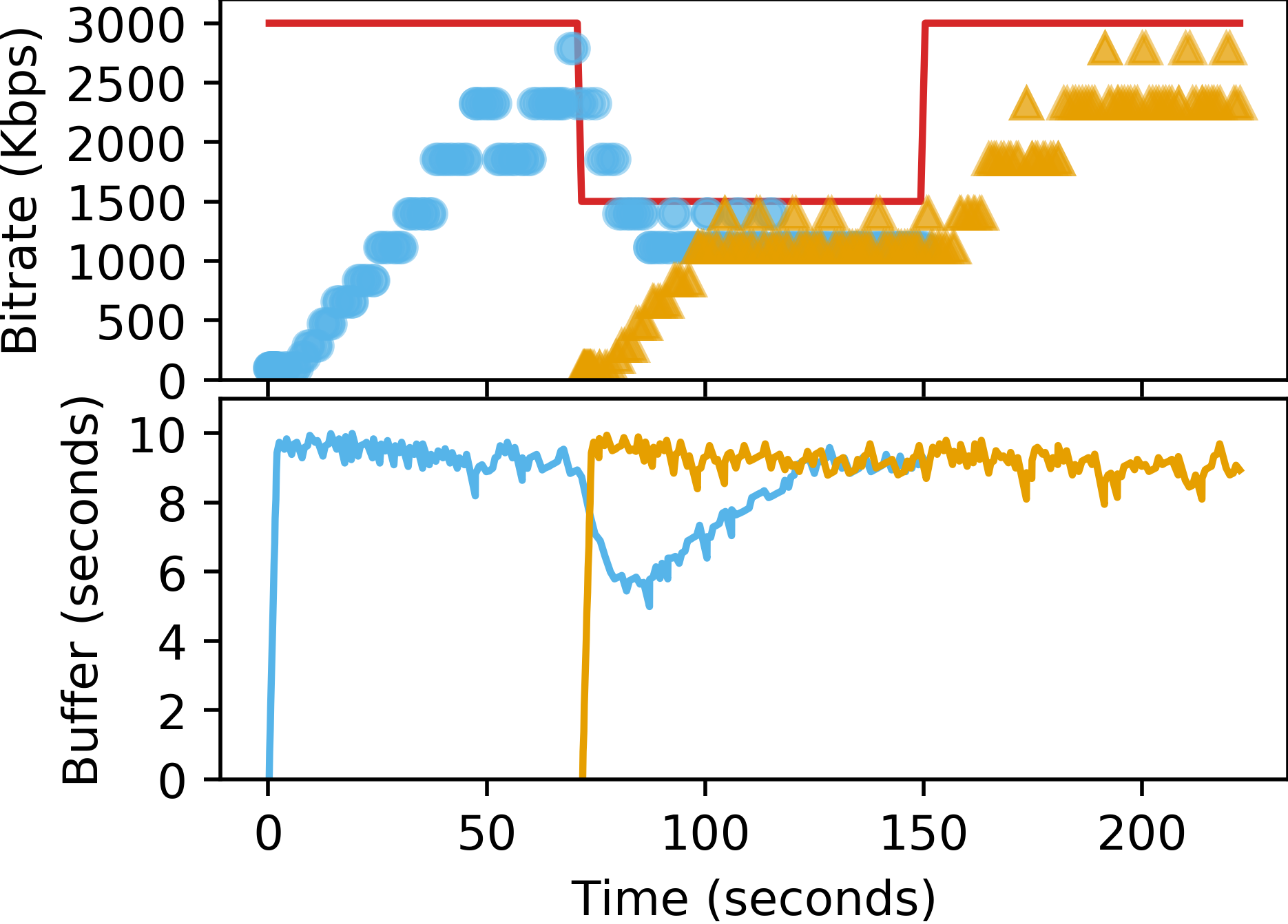}
    \caption{Reactive}
    \label{fig:4_bitrate_1-1_re}
  \end{subfigure}
  
  \vspace{6pt}
  
  \begin{subfigure}{0.47\linewidth}
    \centering
    \includegraphics[width=\textwidth]{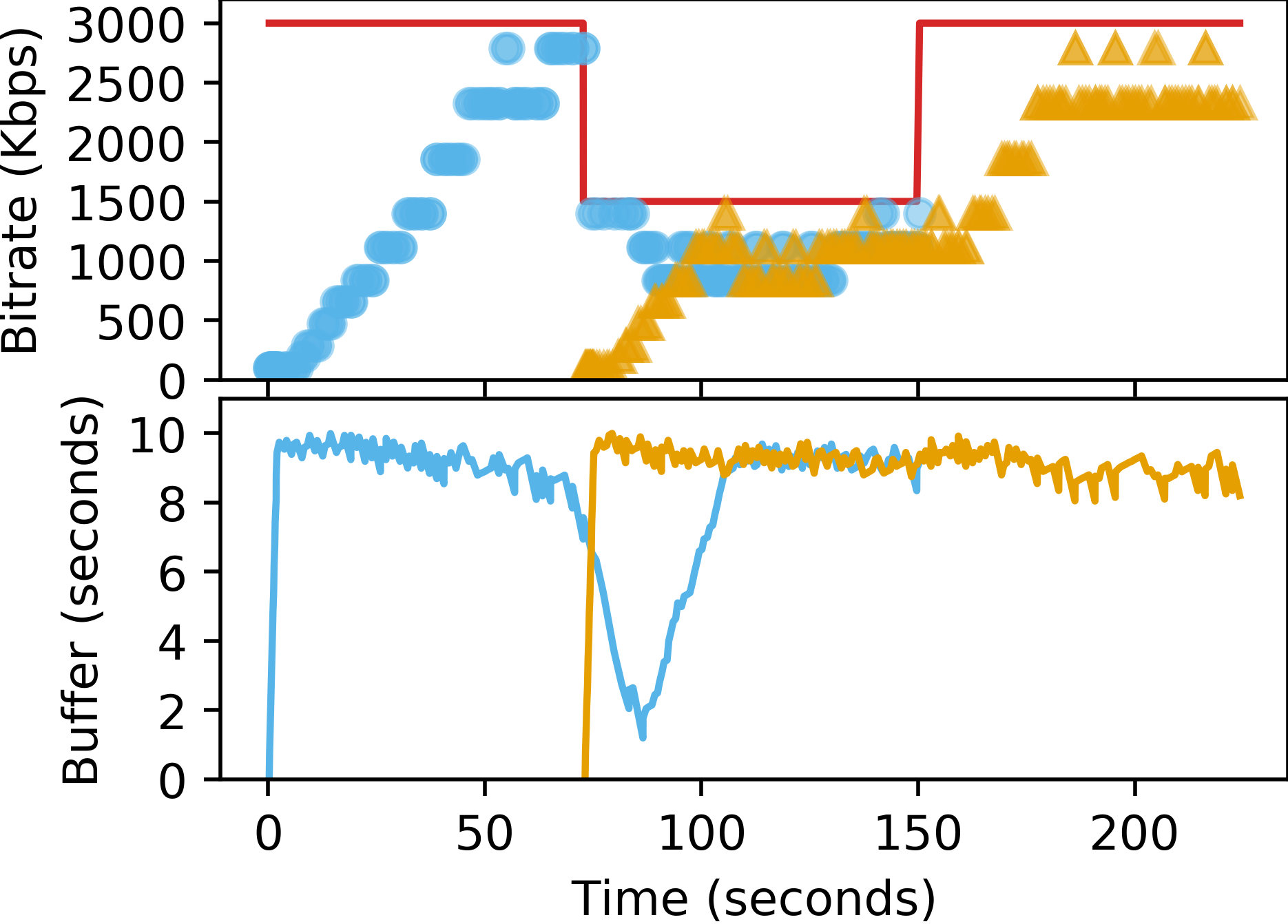}
    \caption{Proactive}
    \label{fig:4_bitrate_1-1_proact}
  \end{subfigure}
  \hfill
  \begin{subfigure}{0.47\linewidth}
    \centering
    \includegraphics[width=\textwidth]{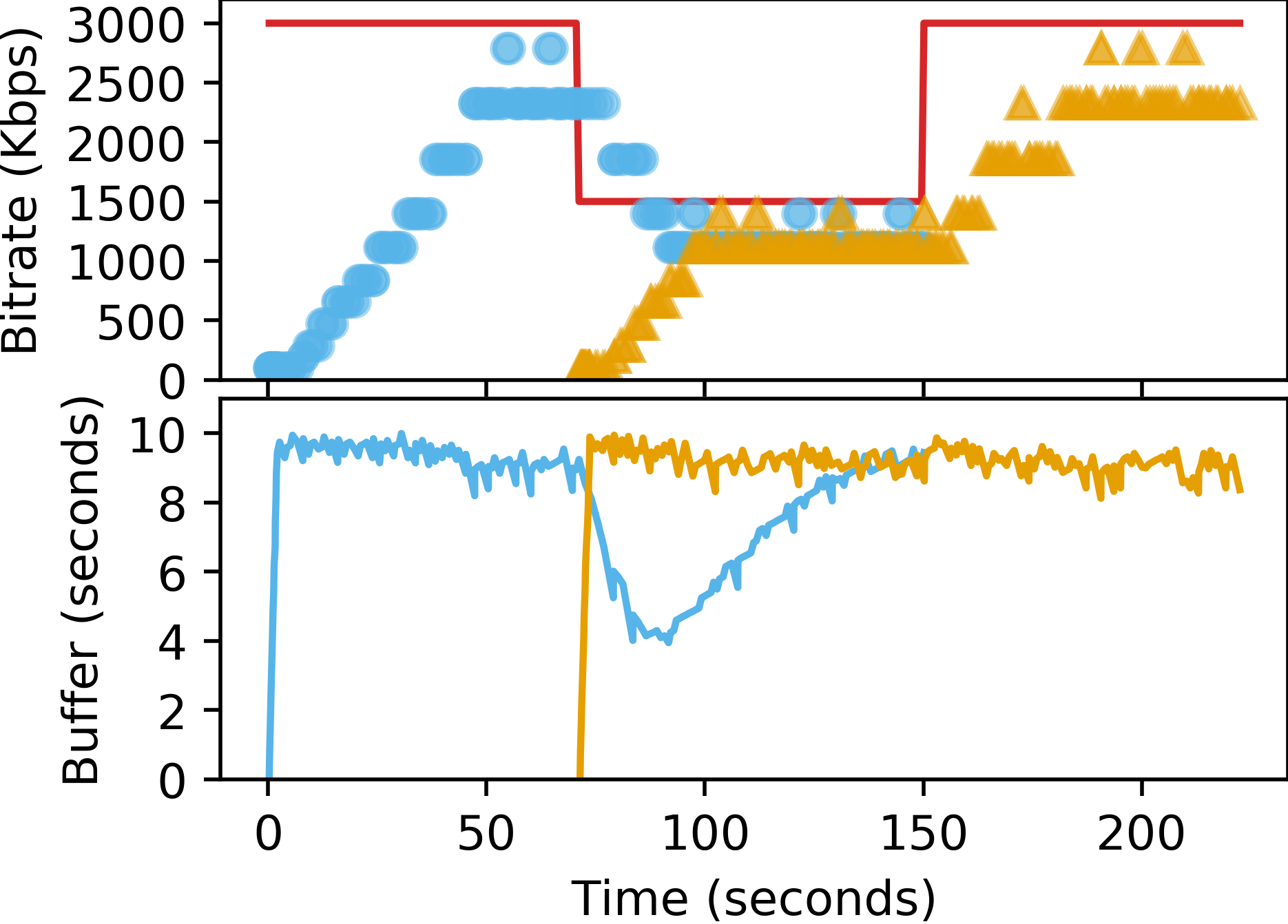}
    \caption{Proposal}
    \label{fig:4_bitrate_1-1_proposal}
  \end{subfigure}
  
  \vspace{8pt}
  
  \begin{subfigure}{0.9\linewidth}
    \centering
    \includegraphics[width=0.5\textwidth]{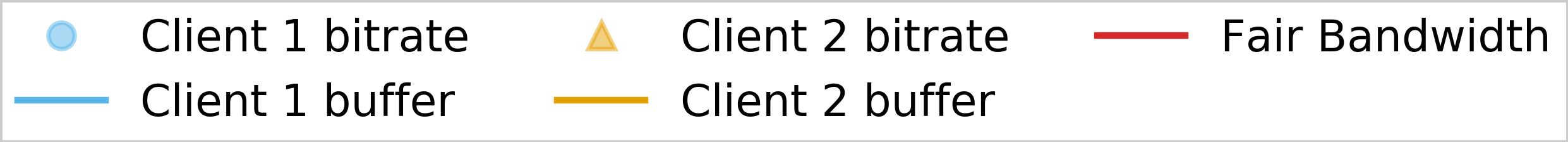}
  \end{subfigure}
  
  \caption{Time-varying performances of all methods in subscenario 2-A}
  \label{fig:4_bitrate_1-1}
\end{figure}

%% file: figures/4_bitrate_1-2.tex
\begin{figure}[ht!]
  \centering
  \begin{subfigure}{0.47\linewidth}
    \centering
    \includegraphics[width=\textwidth]{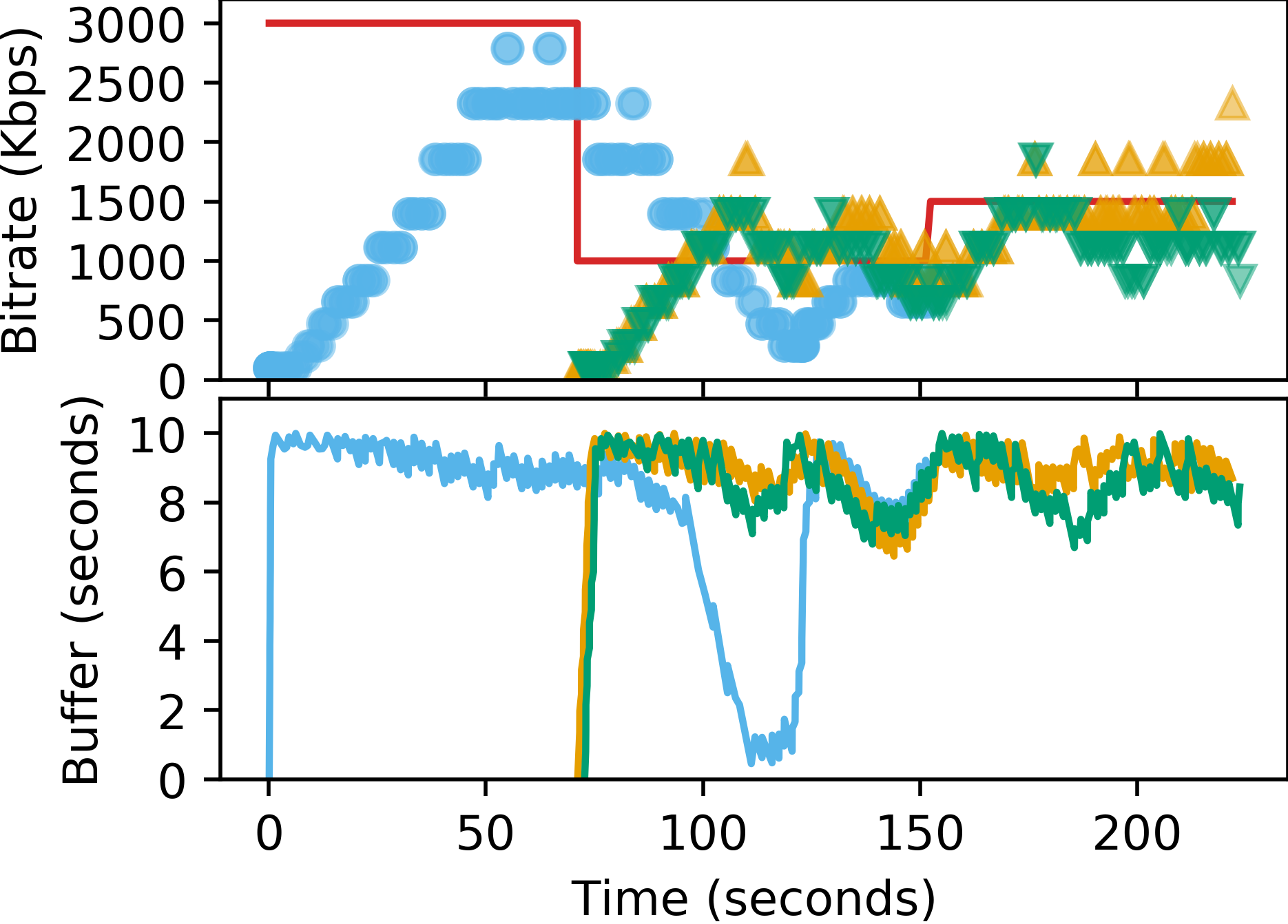}
    \caption{No-Proxy}
    \label{fig:4_bitrate_1-2_no}
  \end{subfigure}
  \hfill
  \begin{subfigure}{0.47\linewidth}
    \centering
    \includegraphics[width=\textwidth]{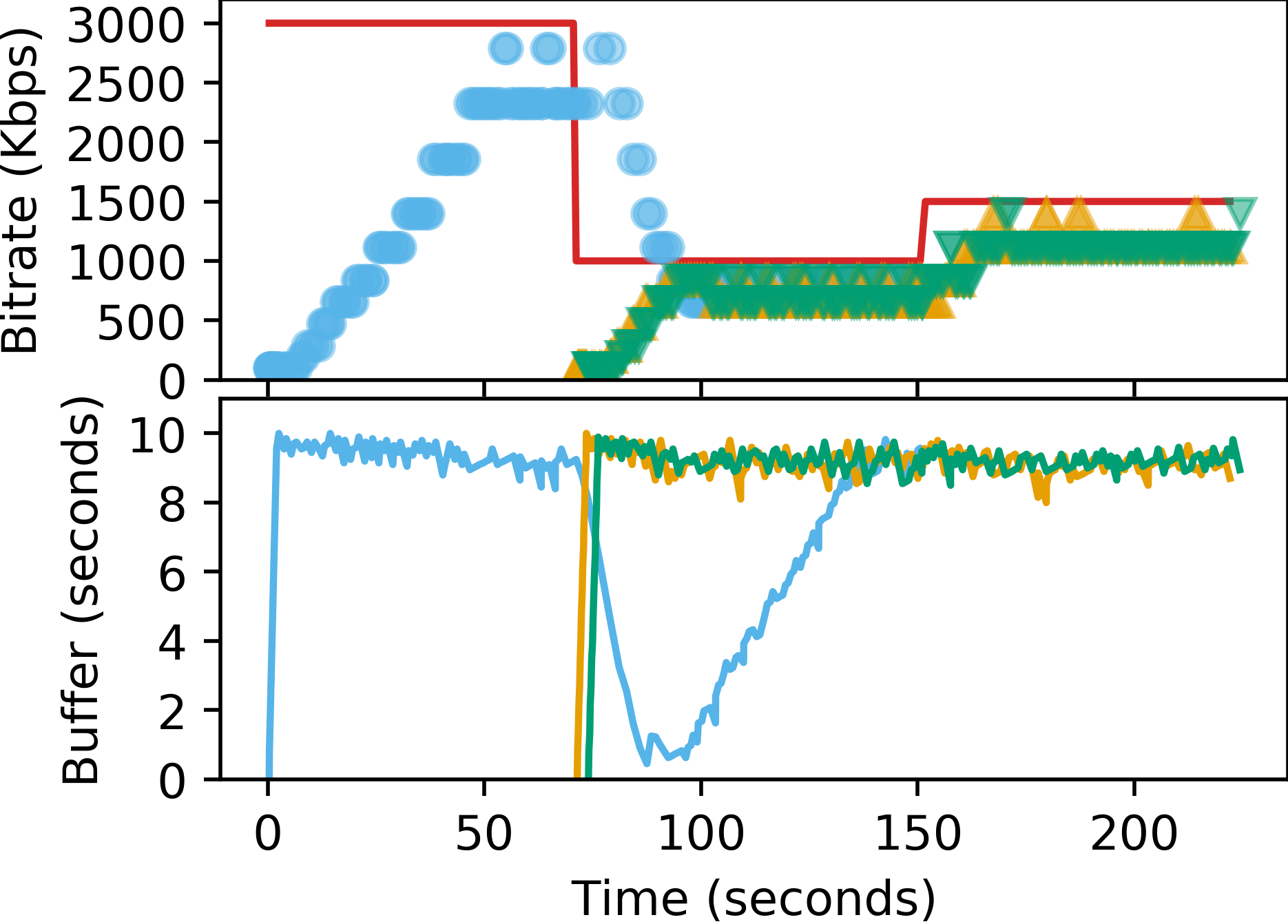}
    \caption{Reactive}
    \label{fig:4_bitrate_1-2_re}
  \end{subfigure}
  
  \vspace{6pt}
  
  \begin{subfigure}{0.47\linewidth}
    \centering
    \includegraphics[width=\textwidth]{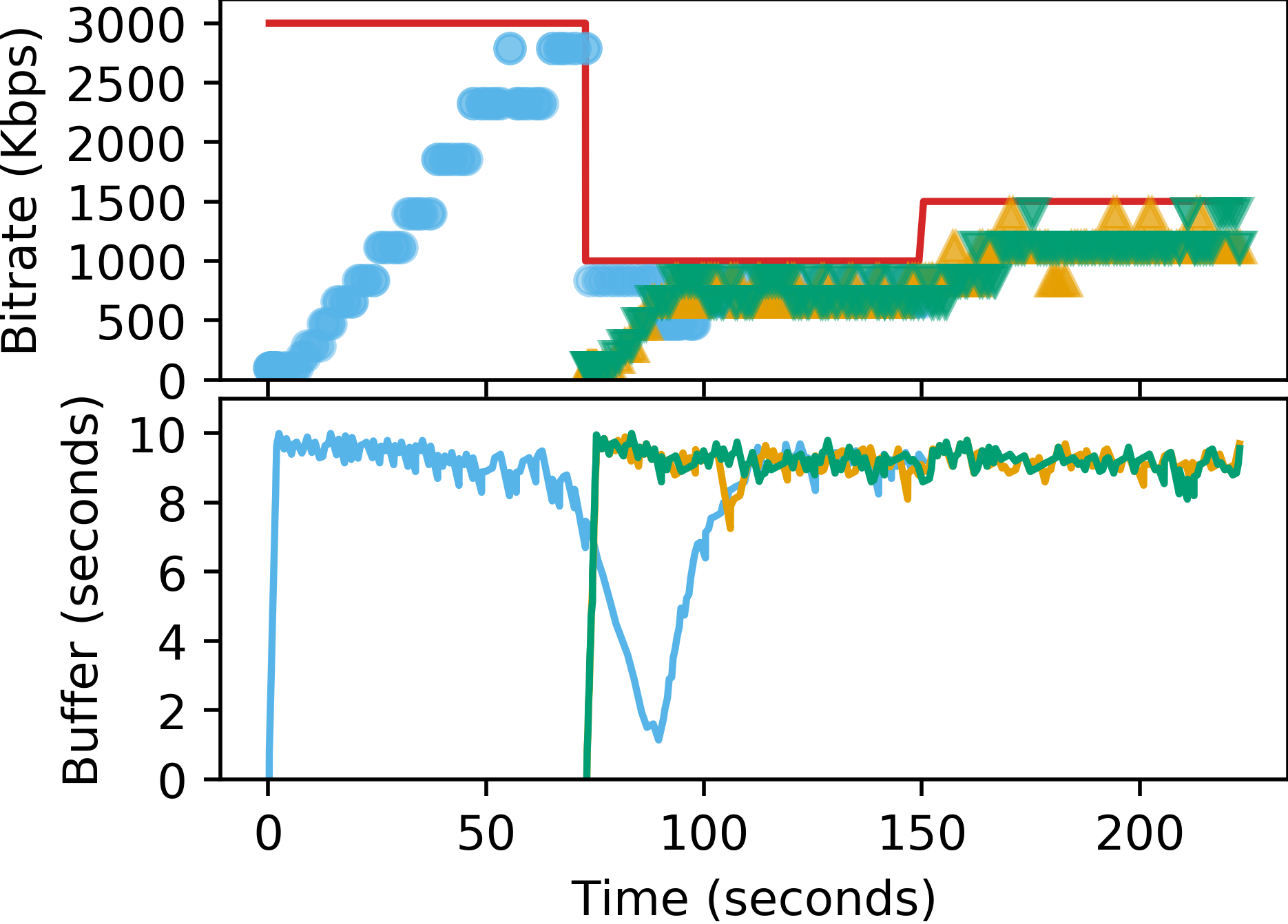}
    \caption{Proactive}
    \label{fig:4_bitrate_1-2_proact}
  \end{subfigure}
  \hfill
  \begin{subfigure}{0.47\linewidth}
    \centering
    \includegraphics[width=\textwidth]{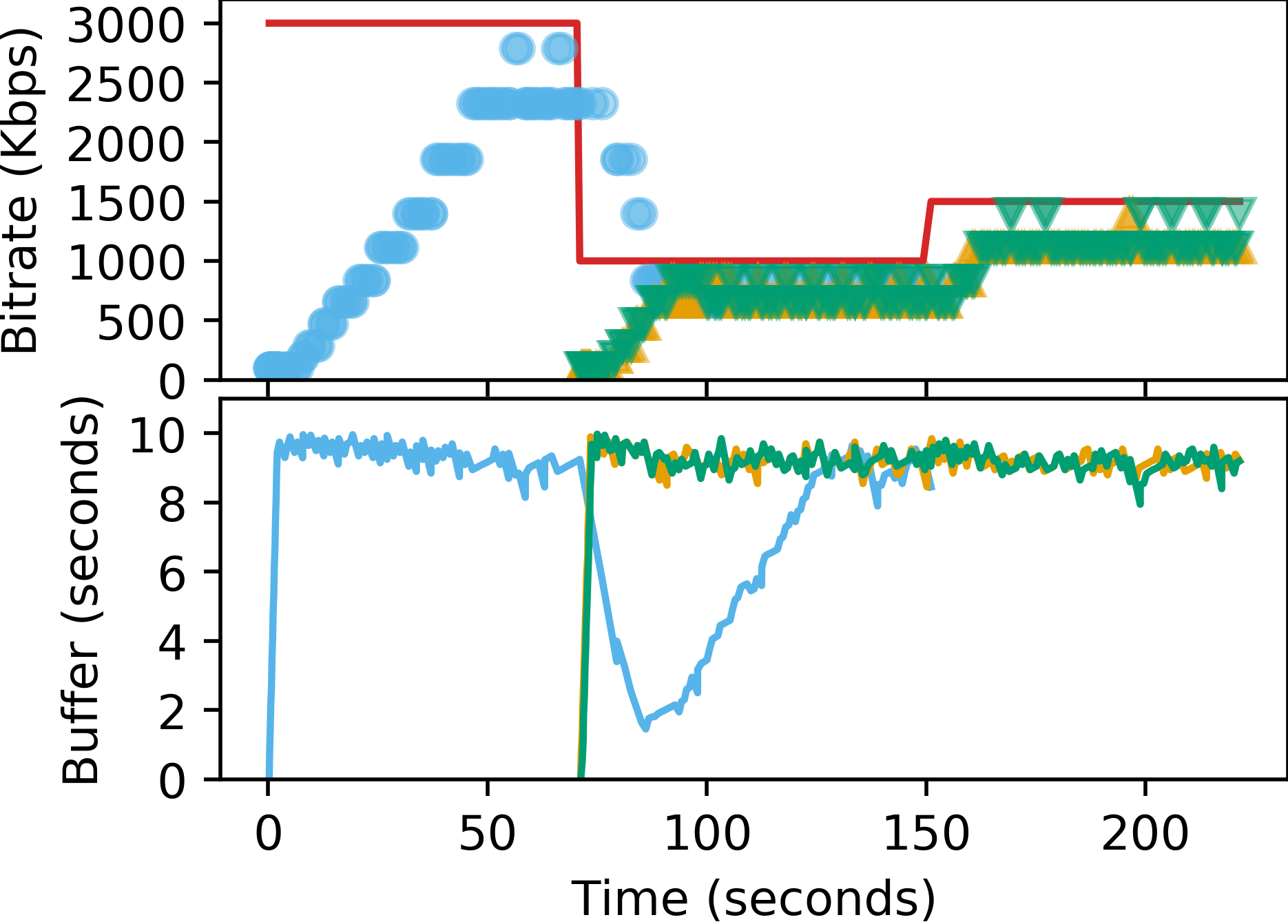}
    \caption{Proposal}
    \label{fig:4_bitrate_1-2_proposal}
  \end{subfigure}
  
  \vspace{8pt}
  
  \begin{subfigure}{0.9\linewidth}
    \centering
    \includegraphics[width=0.7\textwidth]{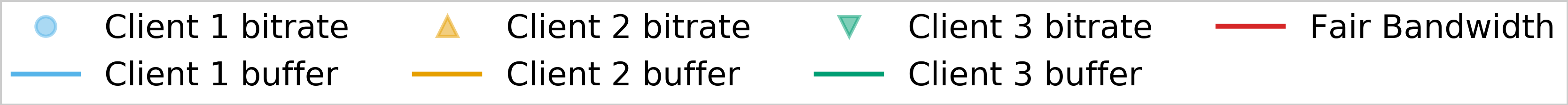}
  \end{subfigure}
  
  \caption{Time-varying performances of all methods in subscenario 2-B}
  \label{fig:4_bitrate_1-2}
\end{figure}

%% file: tables/4_QoERelated.tex
\begin{table}[ht!]
  \centering
  \caption{Adaptation delay and bitrate degradation amplitude of the client $a_1$ when using network-assisted methods in scenario \#2}
  \centering
  \begin{tabular}{ c c m{8em} m{8em} m{8em} }
    \toprule
    \textbf{Subscenario} & \textbf{Metric} & \textbf{Reactive} & \textbf{Proactive} & \proposal{}\\
    \midrule
    \multirow{2}{8em}{\centering \textbf{2-A}} 
    & \textbf{$\Delta{}t$} & 10.001 s & 1.993 s & 9.747 s \\
    & \textbf{$\Delta{}r$} & 454 Kbps & 1016.4 Kbps & 454 Kbps \\ 
    \midrule
    \multirow{2}{8em}{\centering \textbf{2-B}} 
    & \textbf{$\Delta{}t$} & 19.095 s & 2.233 s & 13.547 s \\
    & \textbf{$\Delta{}r$} & 280 Kbps & 1766.2 Kbps & 504.6 Kbps \\ 
    \bottomrule
  \end{tabular}
  \label{tbl:4_QoERelated}
\end{table}

%% file: tables/4_Push.tex

\begin{table}[h]
  \caption{Number of HTTP responses and PUSH\_PROMISE frames of client $a_1$ when using network-assisted methods in scenario \#2}
  \centering
  \begin{tabular}{ ccccc }
    \toprule
    \textbf{Subscenario} & \textbf{Metric} & \textbf{Reactive} & \textbf{Proactive} & \proposal{}\\
    \midrule
    \multirow{2}{8em}{\centering \textbf{2-A}} 
    & $Res$ & 100 & 103 & 100 \\
    & $PP$ & 100 & 103 & 100 \\ 
    \midrule
    \multirow{2}{8em}{\centering \textbf{2-B}} 
    & $Res$ & 100 & 105 & 100 \\
    & $PP$ & 100 & 105 & 100 \\ 
    \bottomrule
  \end{tabular}
  \label{tbl:4_Push}
\end{table}

%% file: sections/5_discussion.tex
\begin{figure}[h!]
  \centering
    \includegraphics[width=1\linewidth]{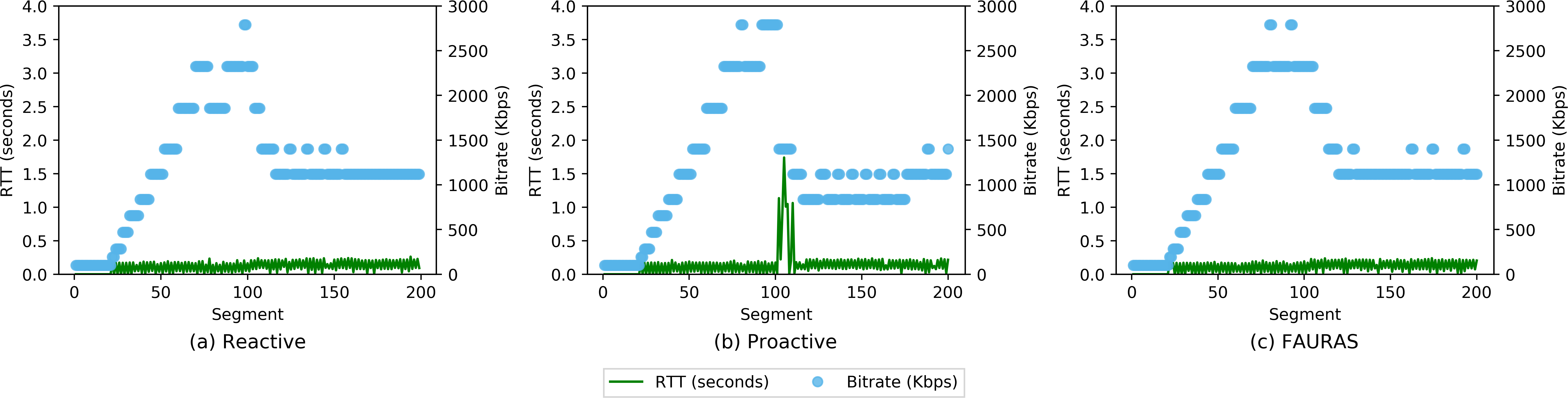}
  \caption{The per-segment bitrate and RTT of the client $a_1$ in subscenario 2-A}
  \label{fig:5_RTT_1-1}
\end{figure}
\begin{figure}[h!]
  \centering
    \includegraphics[width=1\linewidth]{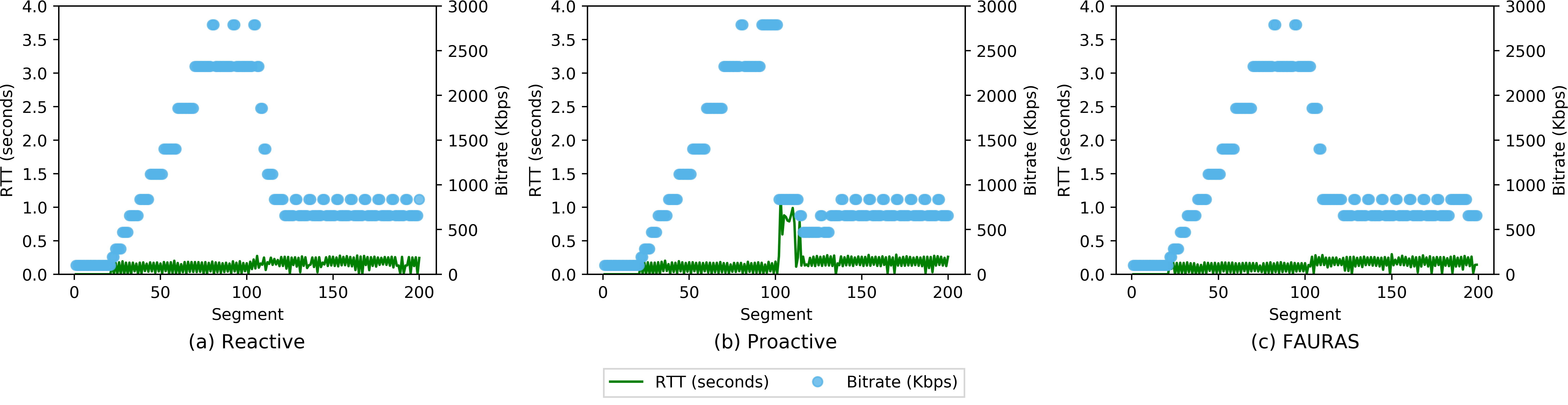}
  \caption{The per-segment bitrate and RTT of the client $a_1$ in subscenario 2-B}
  \label{fig:5_RTT_1-2}
\end{figure}

Based on the evaluation results assessed in Section \ref{section:evaluation}, it has been proven that the proposed \proposal{} significantly improved the fairness of the adaptive streaming over HTTP/2 server push comparing with the \textbf{No-Proxy} method.
Such a performance was identical with the referenced \textbf{Reactive} and \textbf{Proactive} method.
However, although ensuring the fairness, the \textbf{Reactive} method failed to assist the client to react quickly to large changes of the fair bandwidth.
This was because the \textbf{Reactive} method only allocated the bandwidth and did not interfere with the bitrate adaptation of the ABR at the client side.
As a result, even though the gradual quality transition strategy of the ABR was fully respected and the bitrate degradation amplitude was minimized to the smallest, the client failed to maintain its buffer, leading to rebuffering events in subscenario 2-B.
On the other hand, the \textbf{Proactive} method rewrote the bitrate immediately when it exceeded the fair value, thus effectively keeping the buffer undepleted.
Nevertheless, the \textbf{Proactive} method ended up with the highest down-switching amplitude that harmed the user's QoE \cite{Survey_QoE}.
Meanwhile, the \proposal{} not only solved the unfairness problem but also harmonized well the advantages of the above methods.
The bitrate degradation amplitude was significantly reduced comparing to the \textbf{Proactive} method.
Despite that such a bitrate gap was still higher than the \textbf{Reactive} method in subscenario 2-B, the rebuffering events were effectively eliminated.
Moreover, the \textbf{Proactive} method failed to maintain the mechanism of the HTTP/2 server push when it had to overwrite the client's bitrate.
As shown in \ref{subsection:4-3-2_scenario2}, the client discarded the pushed segments with bitrate different from its decision and switched back to the pull-based mechanism of HTTP/1.1.
The per-segment bitrate and RTT of the client $a_1$ in scenario 2-A and 2-B are illustrated on Fig. \ref{fig:5_RTT_1-1} and Fig. \ref{fig:5_RTT_1-2}, respectively, to discussed the consequence of this drawback.
Also, Table \ref{tbl:5_AverageBitrate} summarizes the average bitrate of the client $a_1$ from the time when the new clients joined and caused the fair bandwidth to decrease (segment $100^{th}$) to the time when the streaming session ended (segment $200^{th}$).

\input{tables/5_AverageBitrate}

In both subscenarios, the RTT of the \textbf{Reactive} method (Fig. \ref{fig:5_RTT_1-1}a and \ref{fig:5_RTT_1-2}a) and the \proposal{} (Fig. \ref{fig:5_RTT_1-1}c and \ref{fig:5_RTT_1-2}c) varied within a low stable range.
In contrast, for the case of the \textbf{Proactive} method (Fig. \ref{fig:5_RTT_1-1}b and \ref{fig:5_RTT_1-2}b), the RTT drastically increased for a consecutive number of segments (9 segments for subscenario 2-A and 12 segments for subscenario 2-B) after the new clients joined at segment $100^{th}$.
We speculate that this was because of the deliveries of the wasted PUSH\_PROMISE frames that caused the server to delay the downloads of the responses for those segments \cite{WastePush}.
Since the client performed similar to the HTTP/1.1 during this period, such large RTTs were unavoidable and resulted in bandwidth underutilization \cite{DashHttp2,EvalPushHTTP2}.
Therefore, the client had to lower its bitrate more than the others; the lowest bitrate of the client running the \textbf{Proactive} method was 838 Kbps in subscenario 2-A and 480 Kbps in subscenario 2-B, while those of both the \textbf{Reactive} method and the proposed \proposal{} were 1118 Kbps and 656 Kbps.
As a result, the \textbf{Proactive} method provided the lowest average bitrate for the client $a_1$ in the last 100 segments in both subscenarios as shown in Table \ref{tbl:5_AverageBitrate}.
This underperformance was also because of the overwriting strategy; the bitrate was immediately overwritten once it exceeded the new fair share, thus shortening the time for the client to experience high bitrate levels.
Meanwhile, the \textbf{Reactive} method and our \proposal{} were able to achieve high and relatively similar results.
Comparing with the \textbf{Reactive} method, the proposal showed minor deterioration as the trade-off for preventing the buffer from underflow.







In summary, it is confirmed that our proposed \proposal{} succeeded in improving the fairness in bitrate selection of the clients in adaptive streaming over HTTP/2 server push.
In addition, comparing to the reference methods, the proposal effectively assisted the client to balance the needs of uninterrupted playback and low bitrate degradation amplitude.
Finally, the mechanism of the server push feature was strictly guaranteed that no pushed segment was wastefully discarded.
For those reason, it is fair to conclude that the proposed \proposal{} outperformed other existing methods.


%% file: tables/5_AverageBitrate.tex
\begin{table}[ht!]
  \caption{The average bitrate of the client $a_1$ after the new fair bandwidth is assigned (from segment $100^{th}$ to segment $200^{th}$) when using network-assisted methods in scenario \#2}
  \centering
  \begin{tabular}{ cccc }
    \toprule
    \textbf{Subscenario} & \textbf{Reactive} & \textbf{Proactive} & \proposal{}\\
    \midrule
    \textbf{2-A} & 1268.35 Kbps & 1105.06 Kbps & 1242.29 Kbps \\ 
    \midrule
    \textbf{2-B} & 875.53 Kbps & 721.92 Kbps & 851.11 Kbps \\ 
    \bottomrule
  \end{tabular}
  \label{tbl:5_AverageBitrate}
\end{table}

%% file: sections/6_conclusion.tex
In this paper, a novel proxy-based method is proposed to solve the unfairness in adaptive streaming over HTTP/2 with server push, the \proposal{}.
The proposed method allocates an explicit bandwidth slice for each streaming client and proactively overwrites the bitrate request to ensure the smooth playback.
Through experiments, it has been proven that our proposal not only effectively deals with the unfairness problem but also succeeds in assisting the client to completely avoid rebuffering events and to lower the bitrate degradation amplitude.
Moreover, our method strictly obeys the mechanism of the server push feature, therefore leaving no pushed segment wasted.
For future work, the performance of the \proposal{} will be assessed with multiple bitrate adaptation algorithm.
In addition, the client-oriented characteristics (e.g., subscription plan, device specifications, content preference, etc.) will be considered to investigate the efficiency of the proposed framework in broader scenarios.
